\documentstyle[psfig,epsf,12pt]{ioplppt}           %use this for preprint style

\begin{document}
\title{
Evolution of the superposition of displaced number states with
 the two-atom multiphoton Jaynes-Cummings  model: interference and entanglement
}[Author guidelines for IOP Journals]

\author{ Faisal A. A. El-Orany }
\address{
Department of Mathematics  and Computer Science,
Faculty of Science,
Suez Canal University,
 Ismailia, Egypt}

%\maketitle

\begin{abstract}
In this paper we study the evolution of the two two-level atoms
interacting with a single-mode quantized radiation field, namely,
two-atom multiphoton (, i.e. $k$th-photon) Jaynes-Cummings model
when the radiation field and atoms are initially prepared in the
superposition of displaced number states
 and  excited atomic states, respectively. For
this system we investigate the atomic inversion, Wigner function,
phase distribution  and entanglement. We show that there is a
connection between all these quantities. Moreover, for symmetric
(asymmetric) atoms the system can generate asymmetric (symmetric)
cat states based on the values of the interaction parameters. This
is shown in the behaviors of the Wigner function and phase
distribution. The degree of entanglement for the field-atoms and
the one-atom-remainder tangles depends on the energy follow between
the bipartite. Also it is sensitive to the interference in phase
space and the value of the parameter $k$.

\end{abstract}
%
%  Uncomment out if preprint format required
%
\noindent 

%\noindent
%\maketitle
%
\section{Introduction}

Entanglement is the striking feature of quantum mechanics
revealing the existence of nonlocal correlations among different
parts of a quantum system. A pair of quantum systems is called
entangled if the  measurement on one of them  cannot be performed
independent  of that of the other. Quantum entanglement is a
resource for certain tasks that can be performed faster or in a
more secure way than the classical correlation. For these reasons
entanglement plays an essential role in quantum information, e.g.
quantum computing \cite{prin}, teleportation \cite{tel1},
cryptographic \cite{cry1}, dense coding \cite{dens} and
entanglement swapping \cite{swap}. These new aspects  have
launched intensive experimental efforts to generate entangled
states and theoretical efforts to understand their structures.
 For instance, the entanglement between two qubits in an
arbitrary pure state has been quantified by  the concurrence
\cite{wott}, however, that of the mixed states has been given as
the infimum of the average concurrence over all possible pure
state ensemble decomposition. The closed form for the concurrence
has been expressed by Peres-Horodecki measure \cite{peres}.
Moreover, this measure has been  extended to include a bipartite
system AB, with arbitrary dimensions $D_A$ and $D_B$ in an overall
pure state \cite{rung}. The latter technique has been used to
treat the entanglement for the tripartite quantum system in a
Hilbert space with tensor product structure $2\bigotimes
2\bigotimes \infty$ \cite{tess}, e.g for the two-atom
Tavis-Cummings model in an overall pure state.

The interaction between the radiation field and the two-level
atom, namely, Jaynes-Cummings model (JCM) is an important topic in
quantum optics since it is  solvable in the framework of the
rotating wave approximation (RWA) and it is experimentally
implemented \cite{exp,{boca}}. Also the JCM is a rich source for
the nonclassical effects, e.g. the revival-collapse phenomenon
(RCP) in the evolution of the atomic inversion \cite{eber} and the
generation of the cat states at one-half of the revival time
\cite{cat,{fais}}. The importance of the JCM is increased as a
result of  the progress
 in the quantum information \cite{knig}.
The JCM has been generalized and extended in different directions
\cite{dir}. One of these directions, which is of  particular
interest, is the two two-level atoms  interacting with a single
quantized electromagnetic field (TJCM)
\cite{ros1,{ros2},{ros4},{ros3},{ros5}}. The atomic inversion of
the TJCM  exhibits RCP having forms different from those of the
JCM. This quantity has been investigated when the field is
initially prepared in coherent state
\cite{ros1,{ros2},{ros4},{ros3}}, binomial state \cite{ros5} and
displaced squeezed state \cite{sqz}. Also the entanglement for the
TJCM when the field is initially prepared in coherent state
\cite{tess} and  binomial state \cite{ros5} have been discussed,
too.

The main object in  quantum optics is to increase the nonclassical
effects obtained from the quantum system. This has been achieved
by developing new states beside the standard ones.
 One of these states is the displaced number state
\cite{buz}, which can be generated by
 passing coherent light through a non-linear medium
\cite{buz1}. In the framework of the superposition principle  the
superposition of displaced number states (SDN) has been developed
 as \cite{bas}:
\begin{eqnarray}
\begin{array}{rl}
|\alpha, m\rangle=\lambda_{\epsilon}[\hat{D}(\alpha)+\epsilon
\hat{D}(-\alpha)]|m\rangle\\
\\
=\sum\limits_{n=0}^{\infty}C(n,m)|n\rangle, \label{bin1}
\end{array}
\end{eqnarray}
 where
$\hat{D}(\alpha)$ is the displacement operator, $\epsilon$ is a
real parameter where its value will be specified in the text,
 $\lambda_{\epsilon}$ is the normalization constant having the form:
 \begin{equation}
\lambda^{-2}_{\epsilon}=1+\epsilon^2+ 2\epsilon\exp(-2\alpha^2)
{\rm L}_m(4\alpha^2) \label{binn1}
\end{equation}
and

\begin{equation}
C(n,m)=\lambda_{\epsilon}\langle n| [\hat{D}(\alpha)+\epsilon
\hat{D}(-\alpha)]|m\rangle. \label{bi1}
\end{equation}
The photon-number distribution $P(n)$ related to (\ref{bin1}) is

\begin{equation}
P(n)=|C(n,m)|^2 . \label{biu1}
\end{equation}
 For particular values of the
parameters the state (\ref{bin1}) reduces to  number states,
coherent states, Schr\"{o}dinger-cat states and displaced number
state. The state (\ref{bin1}) can be regarded  as a single-mode
vibration of electromagnetic field suddenly displaced by a
collection of two displacements $\pi$ out of phase with respect to
each other. The generation of (\ref{bin1}) has been established by
the so-called "quantum state engineering" \cite{bas,{eng}}. Quite
recently  the decoherence for  (\ref{bin1}) in the framework of
the standard master equation, which is described by the phase
insensitive attenuators or amplifiers, has been investigated in
\cite{dodn}. Also  the superposition of the $N$ displaced number
states \cite{buz1} and superposition of the squeezed displaced
number states \cite{fais,{faisal1}} have been developed, too.
Finally, the density matrix $\hat{\rho}$ of the superimposed
states  has  two parts, namely, statistical-mixture part
$\hat{\rho}_S$ and interference part $\hat{\rho}_I$. The former
provides information on the original components of the states and
the latter gives the interference in phase space.

In the present paper we study  the SDN against  the two two-level
atoms interacting with the quantized single-mode electromagnetic
field (TJCM).  In the RWA the Hamiltonian
\cite{ros1,{ros2},{ros4},{ros3},{ros5}}, which controls the
system,  takes the form
%%%%%%%%%%%%%%%%%%%%%%%%%%%%%%%%%%%%%%%%%%%%%%%%%%%%%%%%%%%%%%%%%%%%%%%%
\begin{eqnarray}
\begin{array}{lr}
\frac{\hat{H}}{\hbar}=\hat{H}_0+\hat{H}_I,\\
\\
\hat{H}_0= \omega\hat{a}^{\dagger}\hat{a}+
\omega_a(\hat{\sigma}_1^{z}+\hat{\sigma}_2^{z}),\quad
\hat{H}_I=\sum\limits_{j=1}^2 \lambda_j
(\hat{a}^{k}\hat{\sigma}_j^{+} + \hat{a}^{\dagger
k}\hat{\sigma}_j^{-}),
 \label{6}
 \end{array}
\end{eqnarray}
%%%%%%%%%%%%%%%%%%%%%%%%%%%%%%%%%%%%%%%%%%%%%%%%%%%%%%%%%%%%%%%%%%%%%%%%%%%
where $\hat{H}_0$ and $\hat{H}_I$ are the free and interaction
parts of the Hamiltonian, $\hat{\sigma}_j^{\pm}$ and
$\hat{\sigma}_j^{z}$ are the Pauli spin operators of the $j$th
atom; $\hat{a}\quad (\hat{a}^{\dagger})$ is the annihilation
(creation) operator denoting  the cavity mode, $\omega$ and
$\omega_a$ are the frequencies of the cavity mode and the atomic
systems (we consider that the two atoms have the same frequency),
$\lambda_j$ is the atom-field coupling constant of the $j$th atom
and $k$ is the transition parameter. In the paper we mainly deal
with the ratio  $g=\lambda_2/\lambda_1$ and consider two cases,
namely, symmetric and asymmetric according to $g=1$ and $g\neq 1$,
respectively. Furthermore, we assume that $\omega_a=2k\omega$ (,
i.e. the exact resonance case) and the two atoms and field are
initially in the excited atomic states $|+,+\rangle$ and the SDN,
respectively. The atomic ground state is denoted by $|-\rangle$.
Therefore, the dynamical state of the whole system can be
evaluated as:
\begin{eqnarray}
\begin{array}{lr}
|\Psi(T)\rangle=
\sum\limits_{n=0}^{\infty}C(n,m)\left[X_1(T,n,k)|+,+,n\rangle
+X_2(T,n,k)|+,-,n+k\rangle\right.
\\
\\
\left. +X_3(T,n,k)|-,+,n+k\rangle +X_4(T,n,k)|-,-,n+2k\rangle
\right], \label{10}
\end{array}
\end{eqnarray}
where  $T=\lambda_1 t$ is the scaled time. The explicit forms for
the dynamical coefficients $X_j(T,n,k)$ can be found in
\cite{ros1,{ros2},{ros4},{ros3},{ros5}}, however, for analytical
tasks  we provide the forms of these coefficients for $g=1$:

\begin{eqnarray}
\begin{array}{lr}
X_1(T,n,k)=\frac{n!(n+k)!}{[(n+k)!]^2+n!(n+2k)!}\left[
\frac{(n+k)!}{n!}\cos(T\zeta_n)
+\frac{(n+2k)!}{(n+k)!}\right],\\
\\
X_2(T,n,k)=X_3(T,n,k)=-i
\sqrt{\frac{(n+k)!}{n!}}\frac{\sin(T\zeta_n)}{\zeta_n},\\
\\
X_4(T,n,k)=\frac{(n+k)!\sqrt{n!(n+2k)!}}{[(n+k)!]^2+n!(n+2k)!}\left[
\cos(T\zeta_n)-1\right],
 \label{I4}
 \end{array}
\end{eqnarray}
where
\begin{equation}
\zeta_n=\sqrt{2\frac{(n+k)!}{n!} +2\frac{(n+2k)!}{(n+k)!}}.
\label{I4i}
\end{equation}

For this system we investigate the total atomic inversion
$\langle\sigma_{z}(T)\rangle=\frac{1}{2}[
\langle\sigma_1^{z}(T)\rangle+\langle\sigma_2^{z}(T)\rangle$,
Wigner ($W$) function, phase distribution $P(\Theta)$ and
entanglement between different components of the system. The
object of this study is as follows.  There are some well-known
phenomena occurred for the JCM so that we see the possible
occurrence of such phenomena for the TJCM and what would be  the
influence of the interference in phase space on the behavior of
the system. Also we quantify the entanglement between different
components in the system.  More illustratively, for JCM it has
been shown that
 the Schr\"{o}dinger-cat state can be generated at one-half of
the revival time. Also this issue has been discussed only for the
off-resonance TJCM when $(k,g)=(1,1)$ with $\alpha>>1$ in the
framework of $Q$ function \cite{kim}. Here  we investigate this
behavior for the resonance TJCM the most general case, i.e. for
all values of $g$, via $W$ function, which gives better on quantum
system than $Q$ function. In this regard--for the system under
consideration--we show that for the symmetric (asymmetric) case
the asymmetric (symmetric) Schr\"{o}dinger-cat states can be
generated. This behavior likely occurs at the quarter of the
revival time. Such generation of cat state  is confirmed in the
behavior of the $P(\Theta)$. Also we deduce the asymptotic form
for the $W$ function in the strong-intensity regime. Moreover, we
shows that the degree of the entanglement is sensitive to the
values of both
 $k$, $g$, and $\epsilon$. Specifically, the degree of
 entanglement for the asymmetric
case is greater than that of the symmetric one, and the
interference in phase space decreases the degree of entanglement.
Finally, for $k>1$ the evolution of the tangles can be connected
by the behavior of the corresponding atomic inversion.

The paper is organized in the following order: In section 2 we
investigate the evolution of the $\langle\sigma_{z}(T)\rangle$,
$W$ function and phase distribution.
 In section 3 we investigate the  entanglement
 between different components of the system.
  In section 4 we summarize the main results.

%%%%%%%%%%%%%%%%%%%%%%%%%%%%%%%%%%%%%%%%%%%%%%%%%%%%%%%%%%%%%
\section{ Atomic inversion, Wigner function and phase distribution}
%%%%%%%%%%%%%%%%%%%%%%%%%%%%%%%%%%%%%%%%%%%%%%%%%%%%%%%%%%%%%%
One of the basic quantities  related to the JCM is the atomic
inversion, which represents the difference between the population
of the excited and the ground atomic states. The  evolution of the
atomic inversion  is representative by showing RCP. The occurrence
of the RCP in this quantity has a quantum origin since it
indicates the granular structure of the initial field distribution
and also  reveals the atom-field entanglement.
 A lot of efforts have been
done for  observing the RCP experimentally \cite{exp,{boca},
{meu}, {fas6},{fas1}}. On the other hand, for the JCM it has been
shown that there is a connection between the behaviors of the
atomic inversion and both the $W$ function and phase distribution.
In this section we discuss this situation for the TJCM and compare
the obtained results with those of the JCM. Also we investigate
the influence of the interference in phase space on the behavior
of the system.

We start with the total atomic inversion, which can be evaluated
for (\ref{10}) as
%%%%%%%%%%%%%%%%%%%%%%%%%%%%%%%%%%%%%%%%%%%%%%%%%%%%%%%%%%%%%%%
\begin{figure}
  \vspace{0cm}
\centerline{\epsfxsize=5.5cm \epsfbox{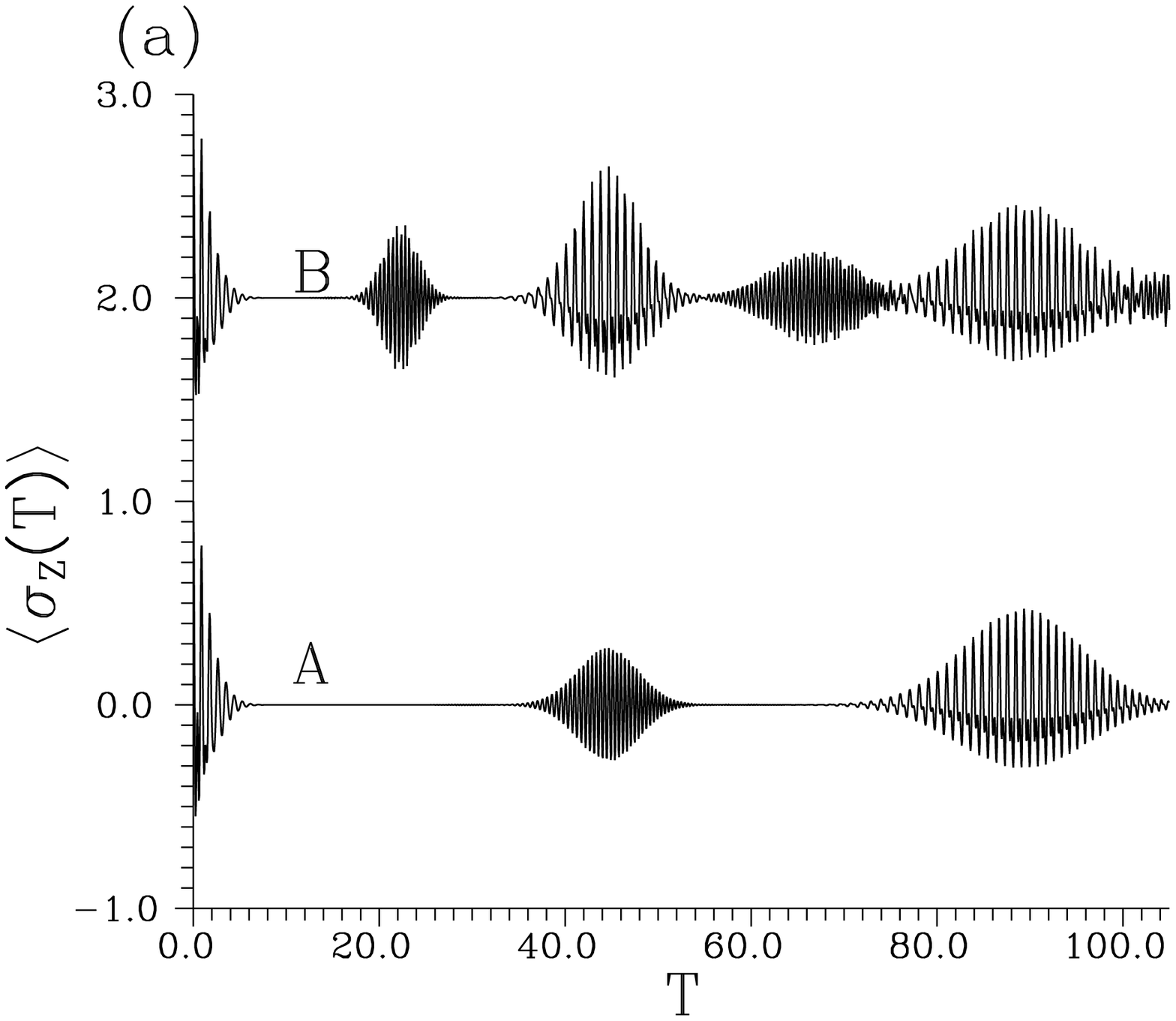}
\\\\\\\\\\\
\epsfxsize=5.5cm \epsfbox{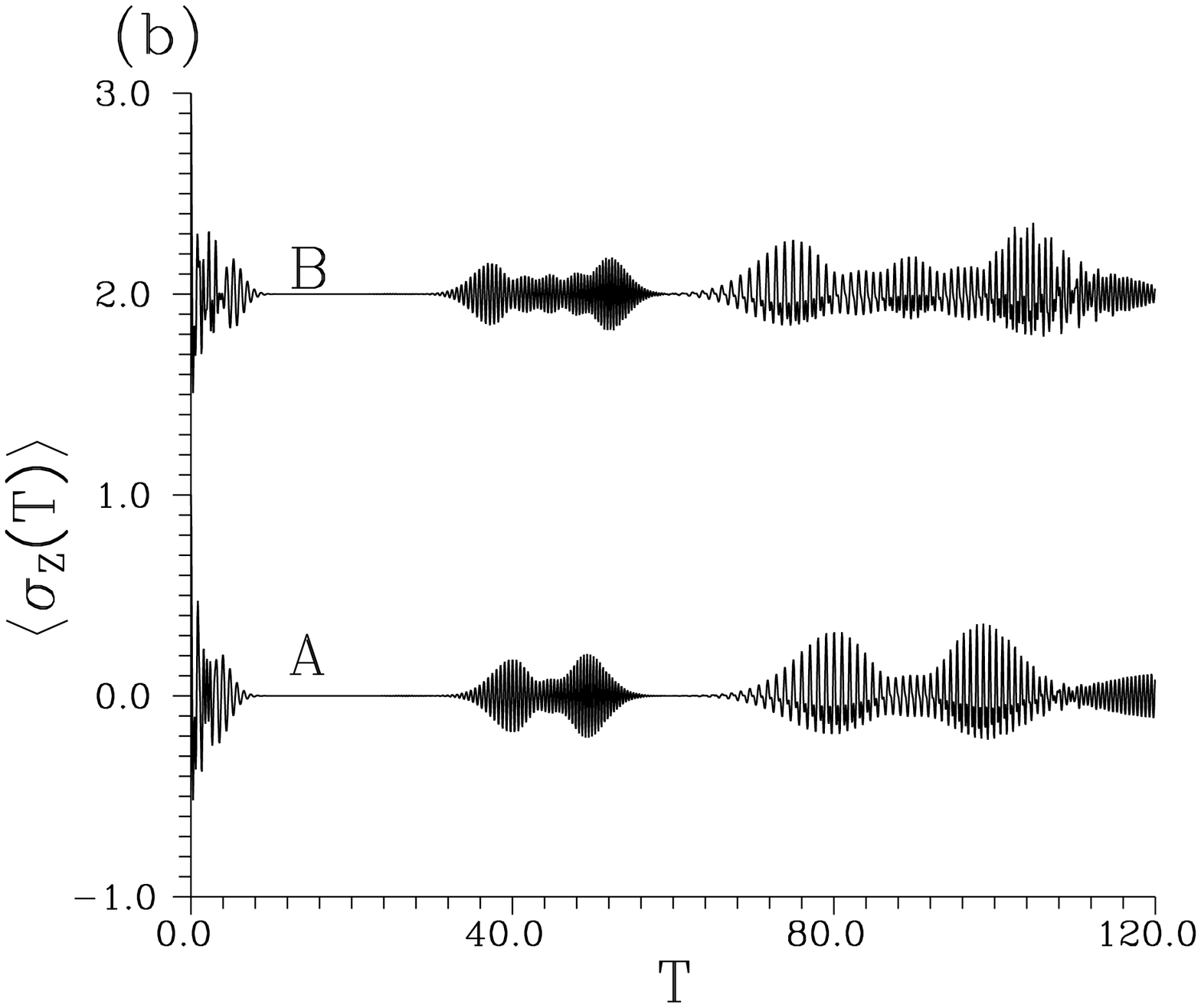}
\\\\\\\\\\\
\epsfxsize=5.5cm \epsfbox{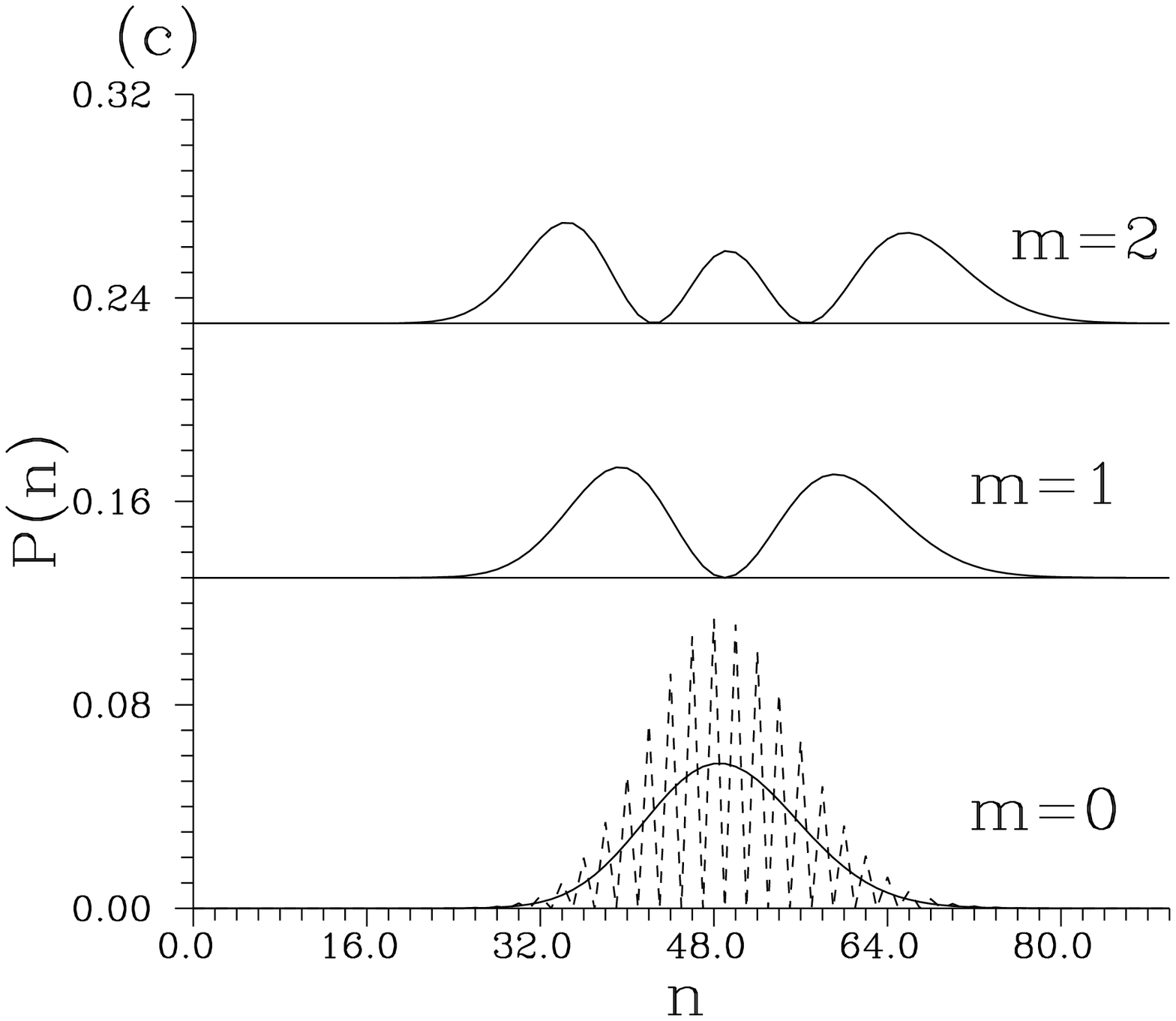} } \vspace{.1cm} \caption{ (a)
and (b) represent the $\langle\sigma_{z}(T)\rangle$ against the
scaled time $T$ for $(\alpha,g)=(7,0.5)$, the curves $A$ and $B$
($\langle\sigma_{z}(T)\rangle+2$) in (a) and (b) are given for
$(\epsilon,m)=(0,0), (1,0)$ and $(0,1),(0,2)$, respectively. (c)
$P(n)$ against $n$ for $(\epsilon,\alpha)=(0,7)$ with the
different values of $m$ as indicated and shifted from bottom by
$0,0.1,0.2$. The dashed curve in (c) represents the $P(n)$ of the
$(\epsilon,\alpha,m)=(1,7,0)$.}
\end{figure}
%%%%%%%%%%%%%%%%%%%%%%%%%%%%%%%%%%%%%%%%%%%%%%%%%%%%%%%%%%%%

\begin{equation}
\langle\sigma_{z}(T)\rangle=
\sum\limits_{n=0}^{\infty}|C(n,m)|^2[|X_1(T,n,k)|^2
-|X_4(T,n,k)|^2].
 \label{13}
\end{equation}
For the symmetric case and in the strong-intensity regime (SR),
i.e. $\alpha>>1$ and $P(n)$ has a smooth envelope. The relation
(\ref{13}) reduces to
 \begin{equation}
\langle\sigma_{z}(T)\rangle=
\sum\limits_{n=0}^{\infty}|C(n,m)|^2\cos
\left(2T\sqrt{n+\frac{3}{2}}\right).
 \label{13a}
\end{equation}
It is evident that  this formula is quite similar to that of the
standard JCM except the factor $3/2$ has to be unity.
 Thus the behavior of the $\langle\sigma_{z}(T)\rangle$ of the TJCM is
quite similar to that of the JCM, i.e. the atomic inversion of the
symmetric TJCM is insensitive of the existence of the other atom.
In this case the revival time  for $(\epsilon,m)= (0,0)$, say, is
$T_r=2\pi\sqrt{\bar{n}+3/2}$.

Now we draw the attention to the $\langle\sigma_{z}(T)\rangle$ of
the asymmetric case (see Figs. 1 for given values of the
interaction parameters). Also we have plotted the corresponding
initial $P(n)$. From this figure one can see that when $P(n)$
exhibits multipeak structure, e.g $m\neq 0$, the  revival patterns
in the corresponding $\langle\sigma_{z}(T)\rangle$ are split. This
is related to that each peak of $P(n)$ provides its own RCP in the
evolution of the $\langle\sigma_{z}(T)\rangle$, which interfere
with those of the others to give such behavior. The comparison
between the curves $A$ and $B$ in Fig. 1(a) shows that the revival
time of the initial cat states is approximately one-half of that
of the initial coherent states and the locations of the symmetric
and asymmetric revival patterns in the two curves are quite
different. This can be easily realized by noting that each
$\hat{\rho}_I$ and $\hat{\rho}_S$ has its own RCP. Furthermore,
the revival time of the former is the two times smaller than that
of the latter.
 Therefore, when the revival patterns
 contributed  by $\hat{\rho}_I$ and $\hat{\rho}_S$ occur at the
 same interaction time asymmetric patterns are exhibited in
 $\langle\sigma_{z}(T)\rangle$.
 This behavior is completely
 different from that of the symmetric case, which we have not presented.

As is well known that for the JCM the evolution of the  $W$
function \cite{fais} as well as the $Q$ function
\cite{ris1,{ris2},{ris3},{ris4}} provides information on the RCP
in the $\langle\sigma_{z}(T)\rangle$. More illustratively, as the
interaction is going on, the $W$ function, which is initially
presented by shifted Gaussian peak, splits into two peaks
counter-rotate on a circle in the complex plane of the
distribution  with interference fringes in between. As the
interaction proceeds the two peaks collide at the opposite end of
the circle  having the initial form, they split again, and so
forth.
%%%%%%%%%%%%%%%%%%%%%%%%%%%%%%%%%%%%%%%%%%%%%%%%%%%%%%%%%%%%%%%
\begin{figure}
  \vspace{0cm}
\centerline{\epsfxsize=5.5cm \epsfbox{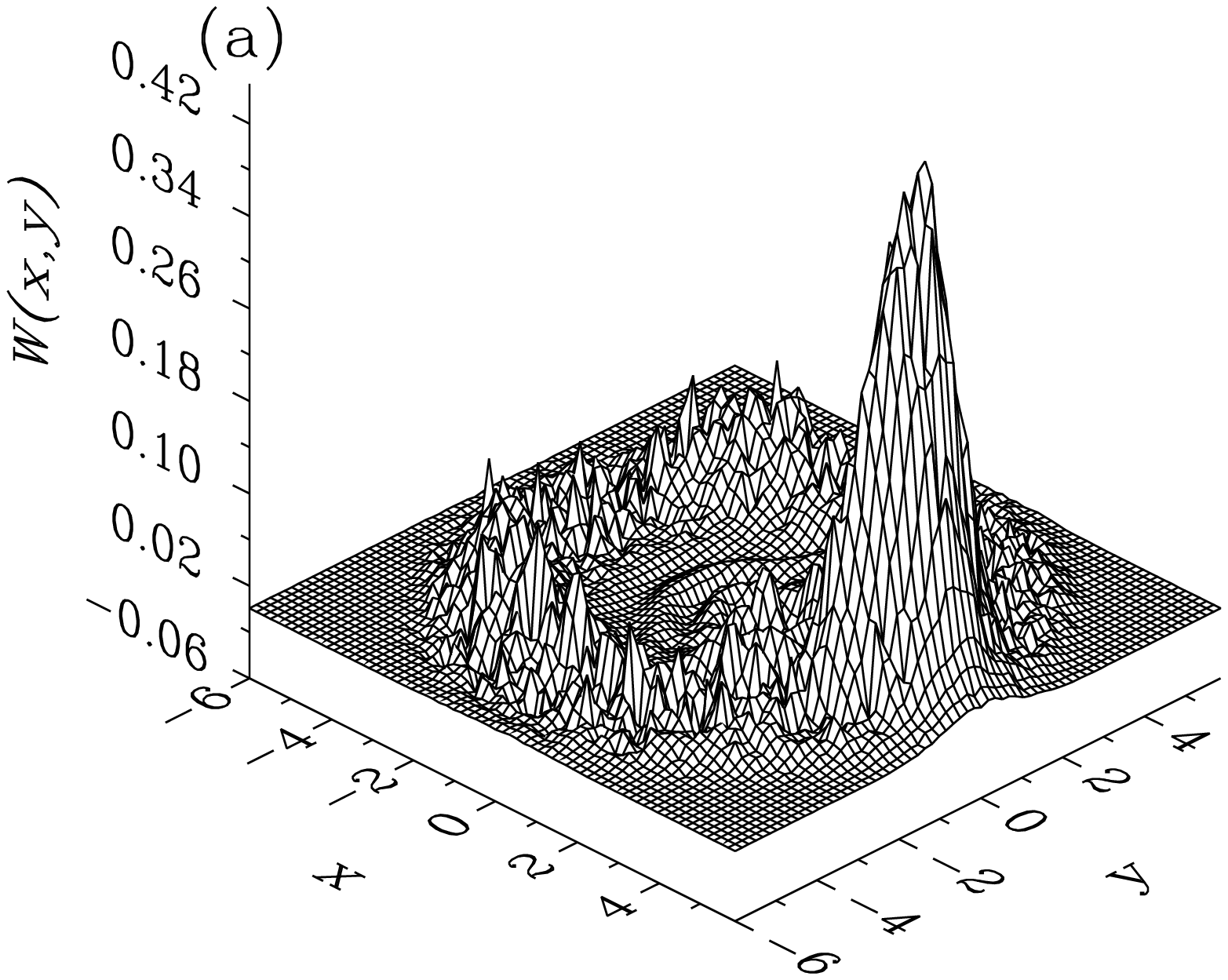}
\\\\\\\\\\\
\epsfxsize=5.5cm \epsfbox{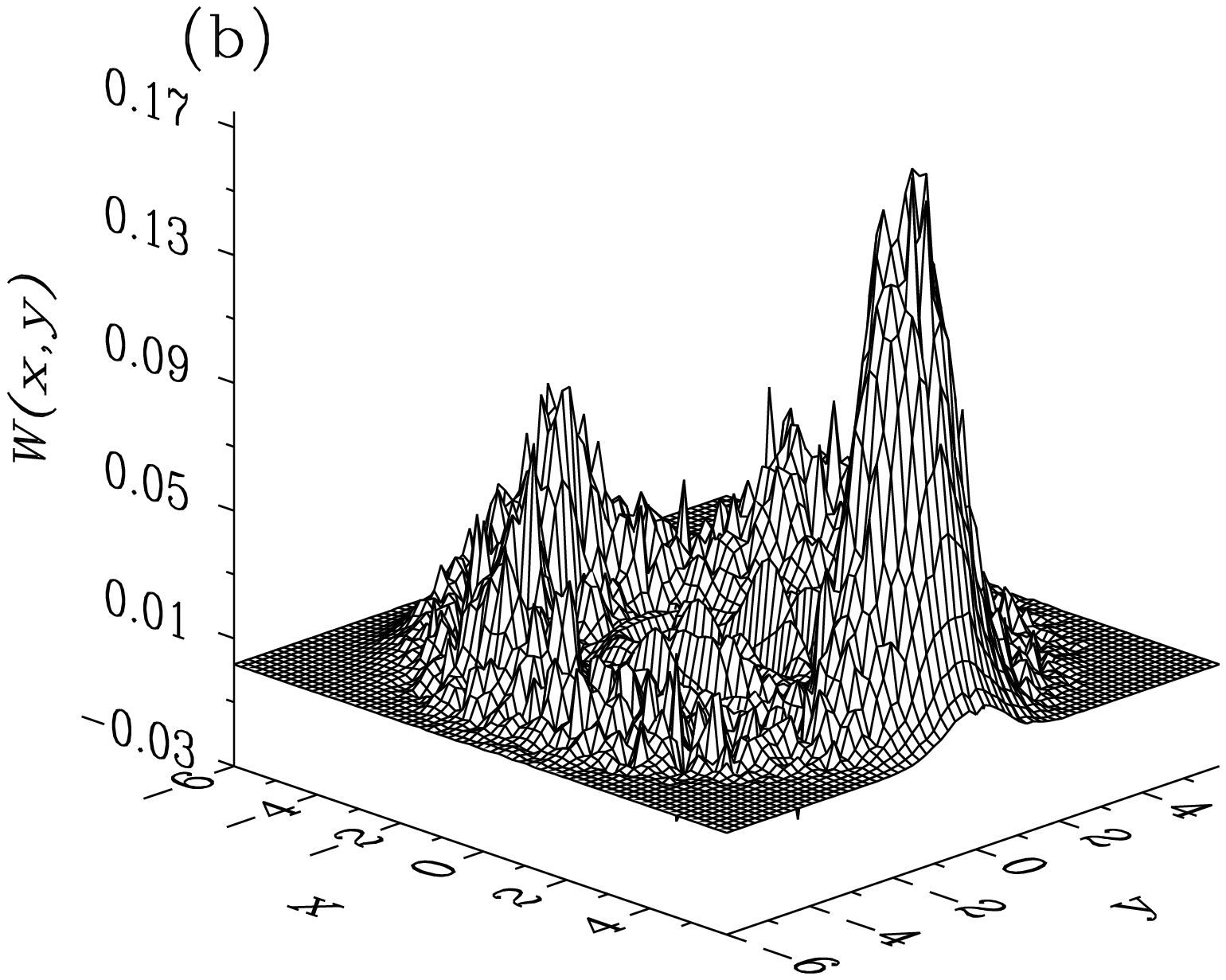}
\\\\\\\\\\\
\epsfxsize=5.5cm \epsfbox{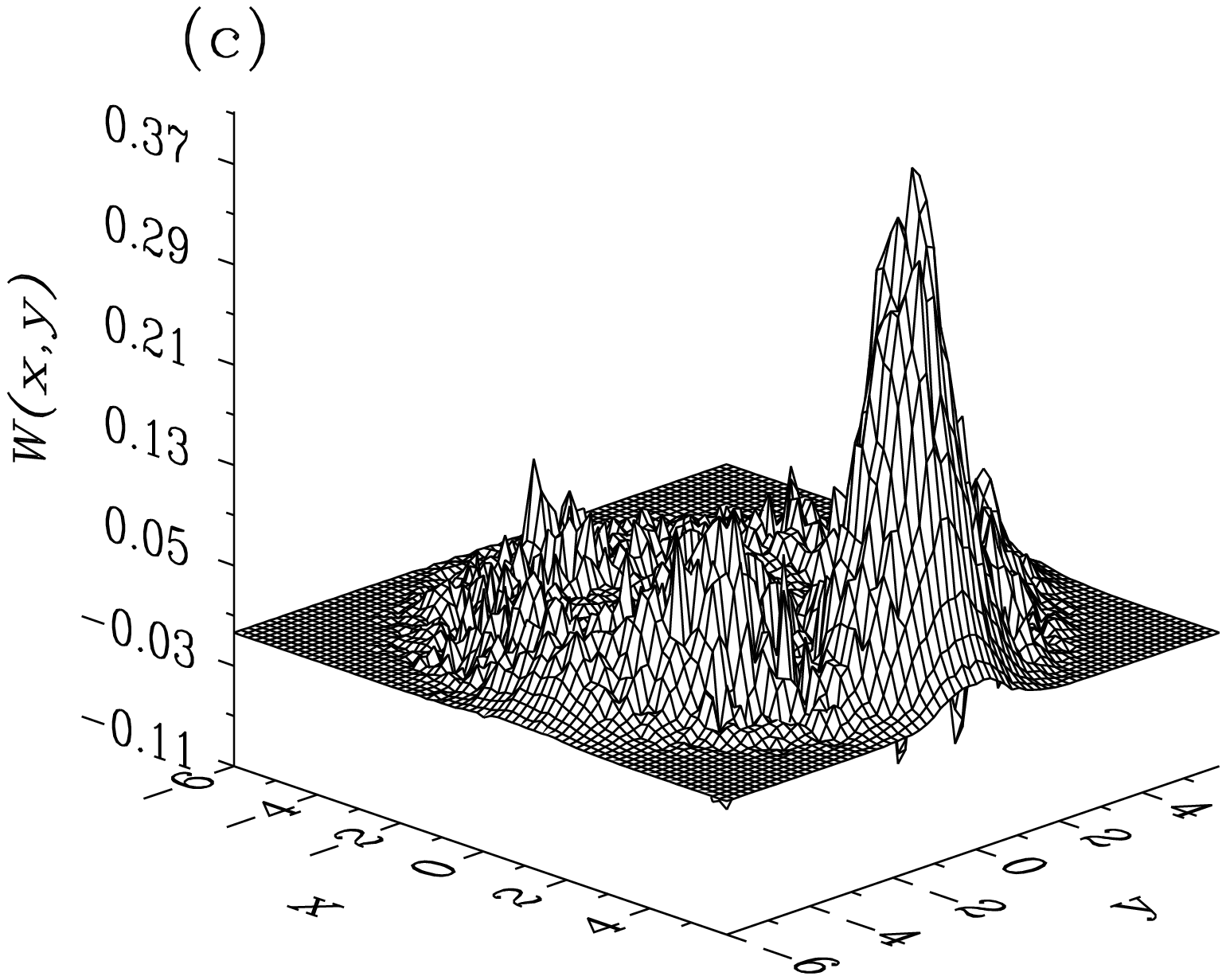} } \vspace{.1cm} \caption{ The
$W$ function against $x$ and $y$ for
$(g,\alpha,\epsilon)=(1,3,0),$ $T=T_r$ (a), $T_r/2$ (b) and
$T_r/4\simeq 5.361749$ (c). $T_r$ refers the revival time.}
\end{figure}
%%%%%%%%%%%%%%%%%%%%%%%%%%%%%%%%%%%%%%%%%%%%%%%%%%%%%%%%%%%%
In the framework of  the $\langle\sigma_{z}(T)\rangle$: the
 collapse regions (revival patterns) in the Rabi oscillations occur in the course of
 the  splitting (collision) of the
distributions of the $W$ function.
 In this regard, the
Schr\"{o}dinger-cat states are generated at the one-half of the
revival time \cite{fais}. Also for the symmetric-off-resonance
TJCM with the initial coherent light  and  using the $Q$ function
it has been numerically shown that asymmetric multipeak structure
can be exhibited for $\alpha>>1$ \cite{kim}. Moreover, the widths
and the heights of these peaks can be controlled by the values of
the detuning parameters and the type of the initial atomic states.
 In this part we investigate the evolution of the $W$ function
 for the TJCM when $\alpha$ is relatively small. This is inspired by the facts
  that
 the $W$ function gives information on the quantum system  better
  than $Q$ function and
 the cat states generated  for small $\alpha$ are more nonclassical than
 those generated for strong initial intensity \cite{fpra}.
 Additionally,
  we will deduce the asymptotic from for the
  state generated by the system when $\alpha>>1$.
  The $W$ function for (\ref{10}) can be easily  evaluated as
\cite{fais}:
\begin{eqnarray}
\fl
\begin{array}{lr}
W(x,y,T) =\frac{\exp(-|z|^2)}{\pi}
\sum\limits_{n,n'=0}^{\infty}C(n,m)C(n',m)
(-1)^{n'}2^{\frac{n-n'}{2}}z^{n-n'}\Bigl\{
X_1(n,T)X_1(n',T)\sqrt{\frac{n'!}{n!}} \\
\\
\times {\rm L}_{n'}^{n-n'}(2|z|^2)+(-1)^k[X_2(n,T)X_2(n',T)+
X_3(n,T)X_3(n',T)]\sqrt{\frac{(n'+k)!}{(n+k)!}} {\rm
L}_{n'+k}^{n-n'}(2|z|^2)\\
\\
+X_4(n,T)X_4(n',T)\sqrt{\frac{(n'+2k)!}{(n+2k)!}} {\rm
L}_{n'+2k}^{n-n'}(2|z|^2)\Bigr\}, \label{14}
\end{array}
\end{eqnarray}
where $z=x+iy$ and $ {\rm L}_{n}^{\nu}(.)$ is the associated
Laguerre polynomial. Generally, we have noted for small values of
$\alpha$ that the entanglement between the two atoms through the
bosonic system leads to that the initial form of the $W$ function
cannot appear again after switching on the interaction.  For
instance, for the symmetric case the single-peak structure coming
from the leading terms (cf. (\ref{I4})) is dominant with
distortion around circle in the $xy$-plane (see Figs. 2). In Figs.
2 the values of the interaction time have been chosen from the
evolution of the corresponding $\langle\sigma_{z}(T)\rangle$.
 Comparison between Figs. 2(a), (b) and (c) shows that the
nonclassicality (, i.e. the negativity) is more pronounced when
$T=T_r/4$. Additionally, the negative values in $W$ function at
$T=T_r$ is much greater than those at $T=T_r/2$, however,
subsidiary peaks start to appear for the latter case. This will be
remarkable in the behavior of the $P(\Theta)$, as we will see
below.

Now we derive the asymptotic form for the $W$ function for the
case $\alpha>>1$ and $(g,k,\epsilon,m)=(1,1,0,0)$ using SR. In
this case we can replace $C(n,0)$ by $C(n+1,0)$ and $C(n+2,0)$ for
the second and third terms in the curely brackets in (\ref{14}),
respectively, and then apply the SR for the arguments of the
trigonometric functions. After some tricks, using the generation
function of the Laguerre polynomial and the Taylor expansion, the
expression (\ref{14}) can be evaluated as
%%%%%%%%%%%%%%%%%%%%%%%%%%%%%%%%%%%%%%%%%%%%%%%%%%%%%%%%%%%%%%%
\begin{figure}
  \vspace{0cm}
\centerline{\epsfxsize=6cm \epsfbox{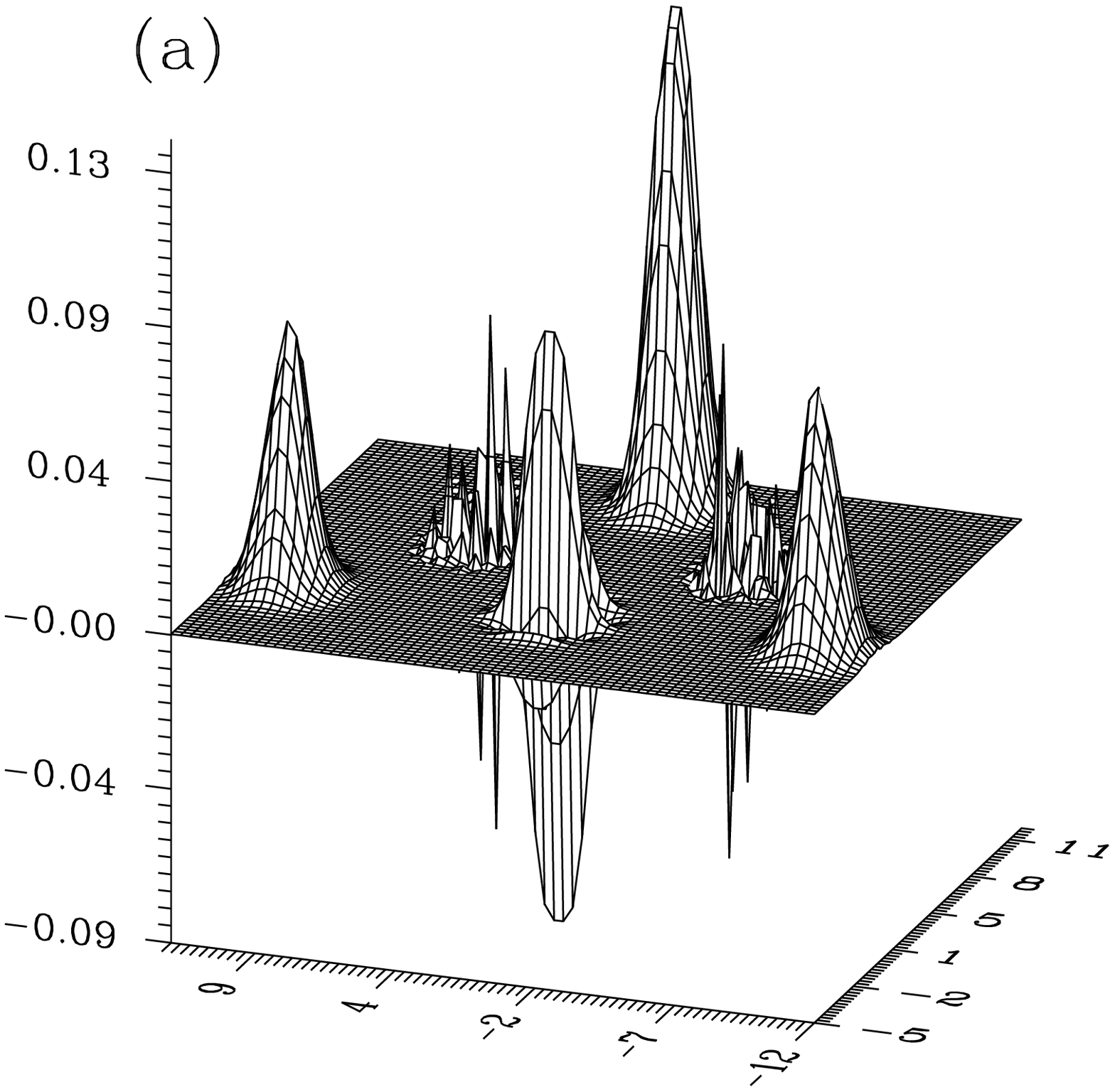}
\\\\\\\\\\\
\epsfxsize=6cm \epsfbox{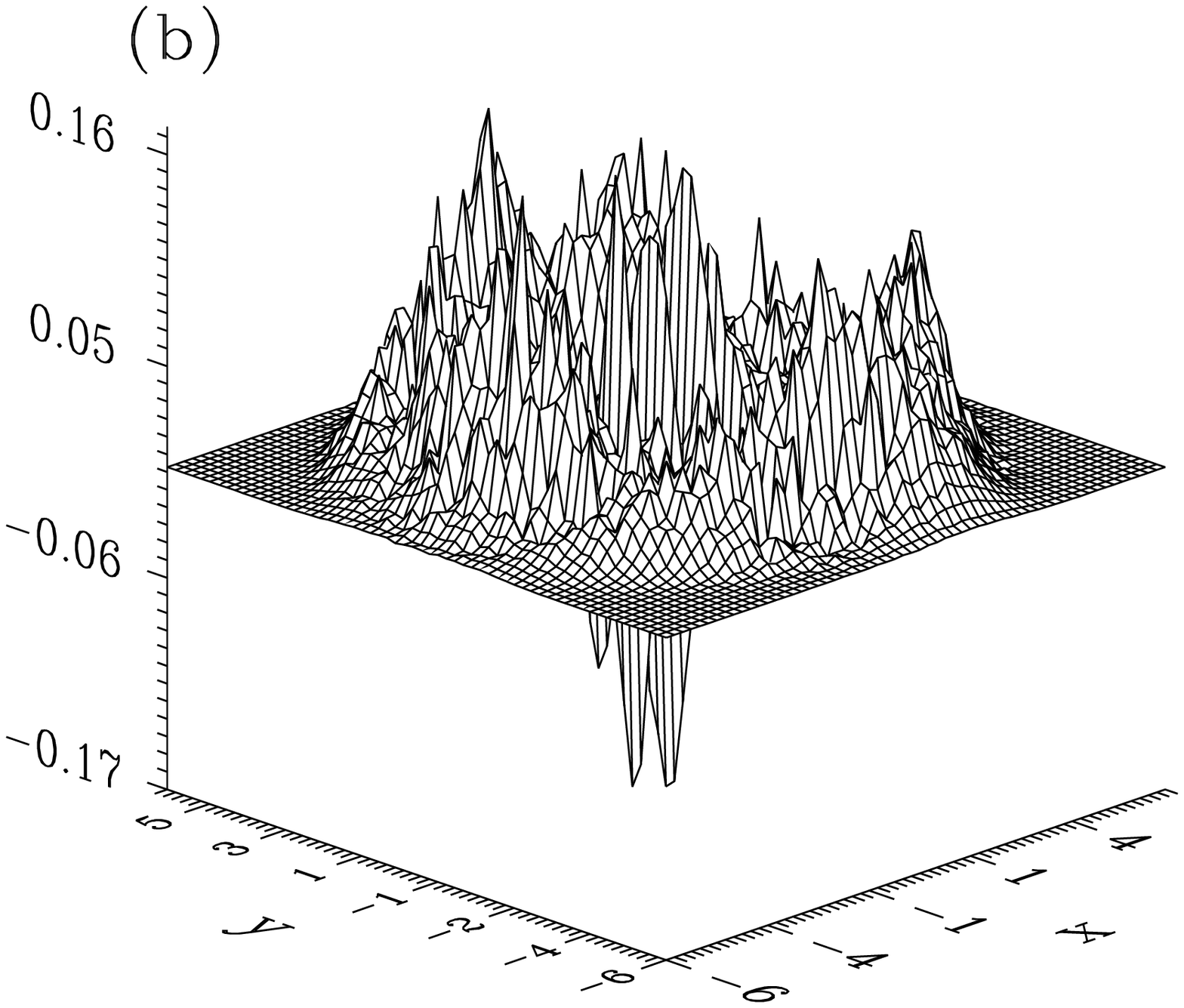}
\\\\\\\\\\\
\epsfxsize=6cm \epsfbox{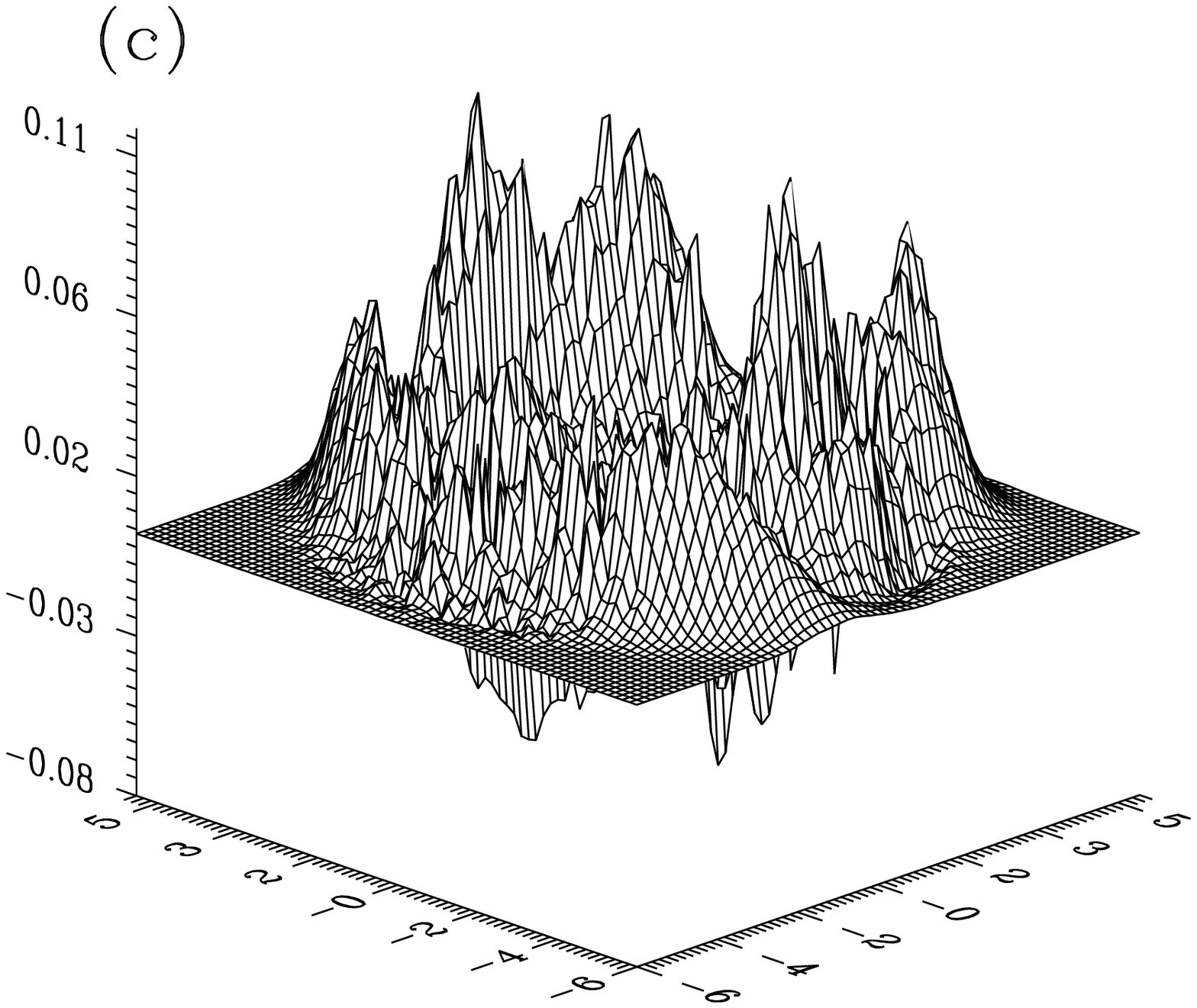} }
 \vspace{.1cm} \caption{ The $W$ function against $x$ and $y$ for the asymptotic
 form (12) $(\alpha,T)=(7,T_r/4)$ (12) (a); the exact form (11)
for $(g,\alpha,\epsilon,T)=(0.5,3,0,19.47003)$ (b) and
$(0.5,3,0,10.18)$ (c).}
\end{figure}
%%%%%%%%%%%%%%%%%%%%%%%%%%%%%%%%%%%%%%%%%%%%%%%%%%%%%%%%%%%%

\begin{eqnarray}
\fl
\begin{array}{lr}
W(x,y,T) \simeq\frac{1}{4\pi}
\Bigl\{2\exp[-y^2-(x-\sqrt{2}\alpha)^2]
+\exp[-(x-\sqrt{2}\alpha\cos\eta)^2-(y-\sqrt{2}\alpha\sin\eta)^2]\\
\\
+\exp[-(x-\sqrt{2}\alpha\cos\eta)^2-(y+\sqrt{2}\alpha\sin\eta)^2]
-2I_{int}(T)\sin\eta \Bigr\}, \label{14w}
\end{array}
\end{eqnarray}
where $\eta=\frac{T}{\sqrt{\bar{n}}}$, the interference part
$I_{int}(T)$ takes the form

\begin{eqnarray}
\fl
\begin{array}{lr}
I_{int}(T) = \exp[-(x-\sqrt{2}\alpha\cos\eta)^2-y^2]\cos[2\eta'
+2\sqrt{2}x\alpha\sin\eta-\alpha^2\sin(2\eta)]\sin\eta\\
\\
+\exp[-(x-\sqrt{2}\alpha\cos^2\frac{\eta}{2})^2]\Bigl\{
\exp[-(y+\frac{\alpha}{\sqrt{2}}\sin\eta)^2]\sin\mu_{+}\\
\\
\exp[-(y-\frac{\alpha}{\sqrt{2}}\sin\eta)^2]\sin\mu_{-}
 \Bigr\} \label{14ww}
\end{array}
\end{eqnarray}
and
\begin{equation}
\fl
 \mu_{\pm} = \eta'-\alpha^2\sin\eta+\sqrt{2}\alpha(x\sin\eta\pm
y\cos\eta)\pm \sqrt{2}\alpha y,  \quad \eta'=T\sqrt{\bar{n}}+
\frac{T}{2\sqrt{\bar{n}}}.
 \label{114www}
\end{equation}
The formula (\ref{14w}) can  give information on the $W$ function
for large values of $\alpha$ for which it is difficult to deal
with the exact form (\ref{14}) because of the associated Laguerre
polynomial.  From (\ref{14w}) it is obvious that the first term is
time independent since it is connecting with the leading terms in
(\ref{I4}). Also  $W$ function includes three individual peaks and
three interference fringes each of them locating between two of
these individual peaks. When $\alpha$ is very large $T_r\simeq
2\pi\sqrt{\bar{n}}$ and  we can extract various facts from
(\ref{14w}).  For instance,  the interference part $I_{int}(T)$
has no (maximum) contribution at $T=T_r,T_r/2 \quad (T_r/4)$. To
be more specific, when $T=T_r$ the $W$ function reduces to that of
the initial one, which is Gaussian bell centered at
$(\sqrt{2}\alpha,0)$, whereas at $T=T_r/2$ the initial bell splits
into two typical peaks localized at $(\pm\sqrt{2}\alpha,0)$ and
eventually at $T=T_r/4$ three-peak structure centered at
$(\sqrt{2}\alpha,0),(0,\pm\sqrt{2}\alpha)$ with interference
fringes in between are exhibited, i.e. three-component cat state
is generated (see Fig. 3(a)). Moreover, we have checked the $Q$
function for these values of the interaction times
 with $\alpha>>1$  and obtained  number
of peaks  as those exhibited for the $W$ function. This behavior
of the $W$ function is different from that of the JCM  even though
the evolution of the atomic inversion for the two systems is quite
similar. This can be easily understood by comparing the
generalized Rabi oscillation for the JCM and the TJCM. Actually,
 for the symmetric case the Rabi oscillation of the TJCM
is approximately two times greater than that of the JCM (cf.
(\ref{I4i})). This fact explains why the cat states in the TJCM
are generated at the quarter of the revival time, however, in the
JCM at the one-half of the revival time.  On the other hand, we
have plotted the $W$ function (\ref{14}) for the asymmetric case
in Figs. 3(b) and (c) for the revival and collapse times,
respectively, when $\alpha=3$. From these figures one can observe
that the $W$ function  exhibits symmetric shapes in phase space,
which are in contrast with the the symmetric case  (see Figs. 2).
The nonclassical effects are more pronounced in the course of the
revival time, which also are greater than those occurred for the
symmetric case (compare Figs. 3(b) and (c) as well as Figs. 3
(b)-(c) and Figs. 2). Furthermore, the shapes presented in Figs.
3(b)-(c) indicate that the generated cat states are of the
microscopic type. The $W$ function of the microscopic cat state
exhibits a complicated shape resulting from that the contribution
of the different components of the state are located close to the
phase-space origin. It is worth mentioning that the microscopic
cat state (, i.e. $\alpha$ is small) provides more nonclassical
effects than the macroscopic one \cite{phase} (compare Figs. 3(a)
and (b)).

Now we turn the attention to  the phase distribution $P(\Theta)$.
To investigate the $P(\Theta)$ we use the Pegg-Barnett formulism
\cite{pegg}, which for the density matrix elements
$\hat{\rho}_{n,n'}(T)$ gives
%%%%%%%%%%%%%%%%%%%%%%%%%%%%%%%%%%%%%%%%%%%%%%%%%%%%%%%%%%%%%%%
\begin{figure} \vspace{0cm}
\centerline{\epsfxsize=5.5cm \epsfbox{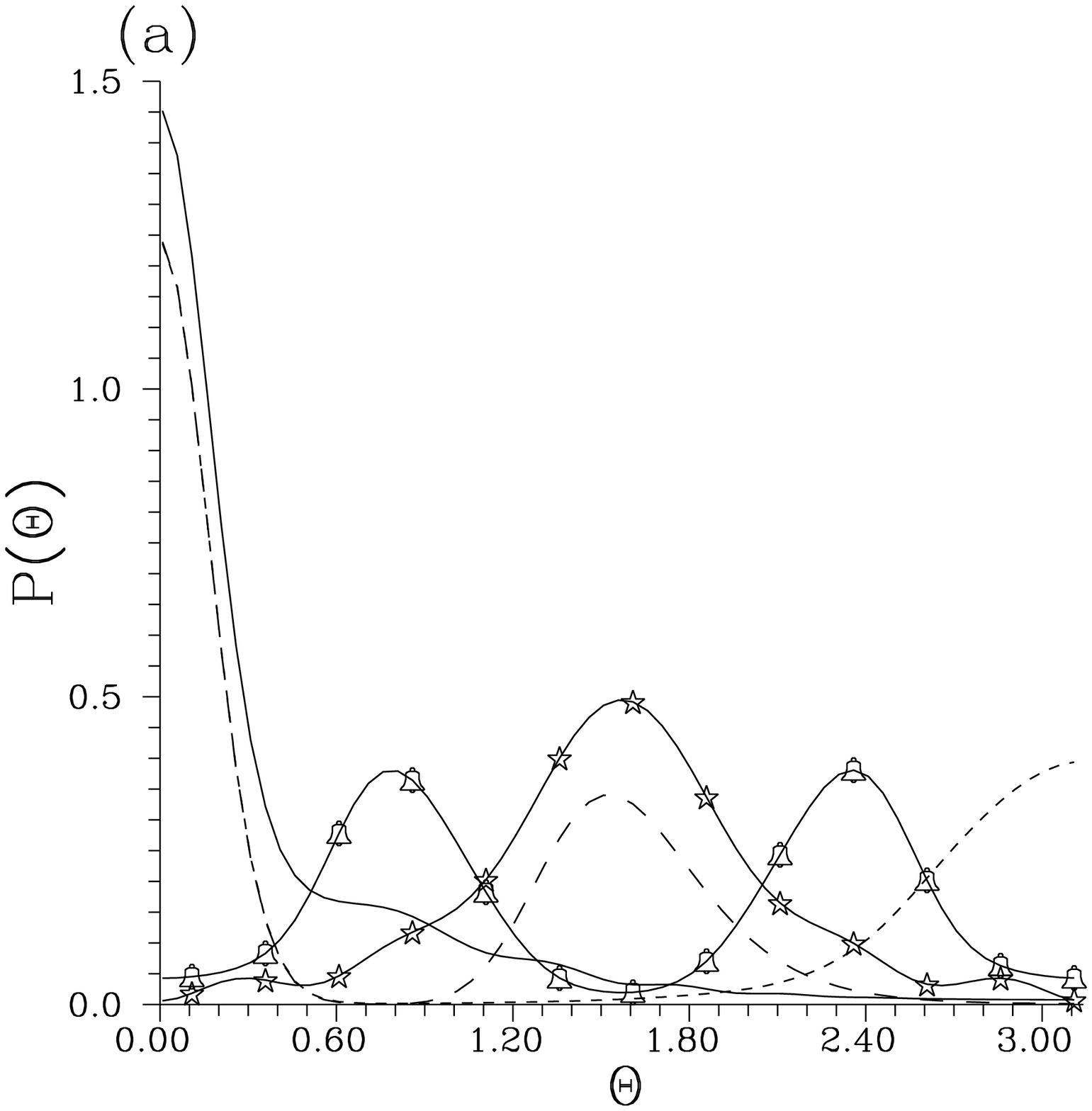}
\\\\\\\\\\\
\epsfxsize=5.5cm \epsfbox{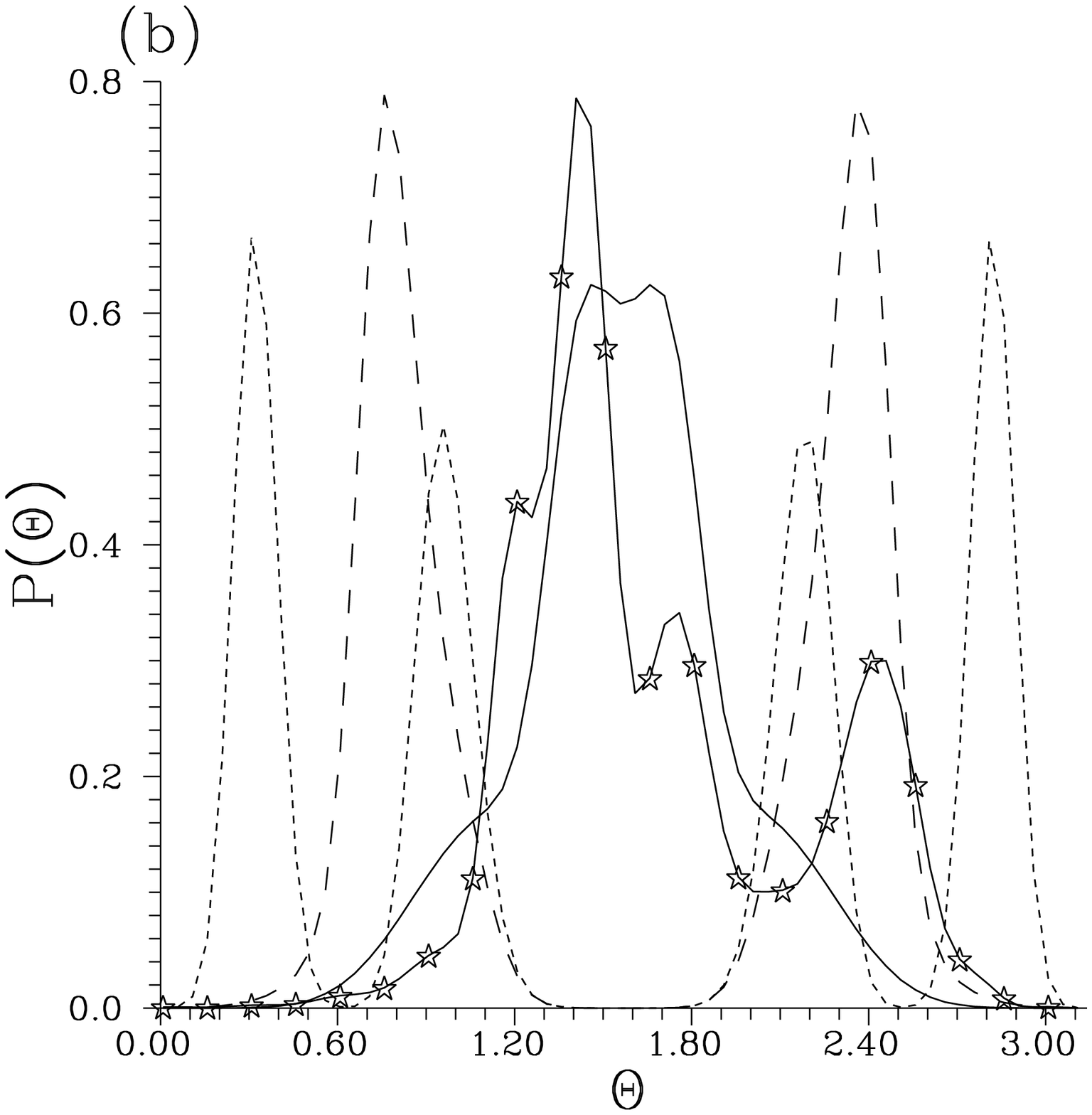}
\\\\\\\\\\\
\epsfxsize=5.5cm \epsfbox{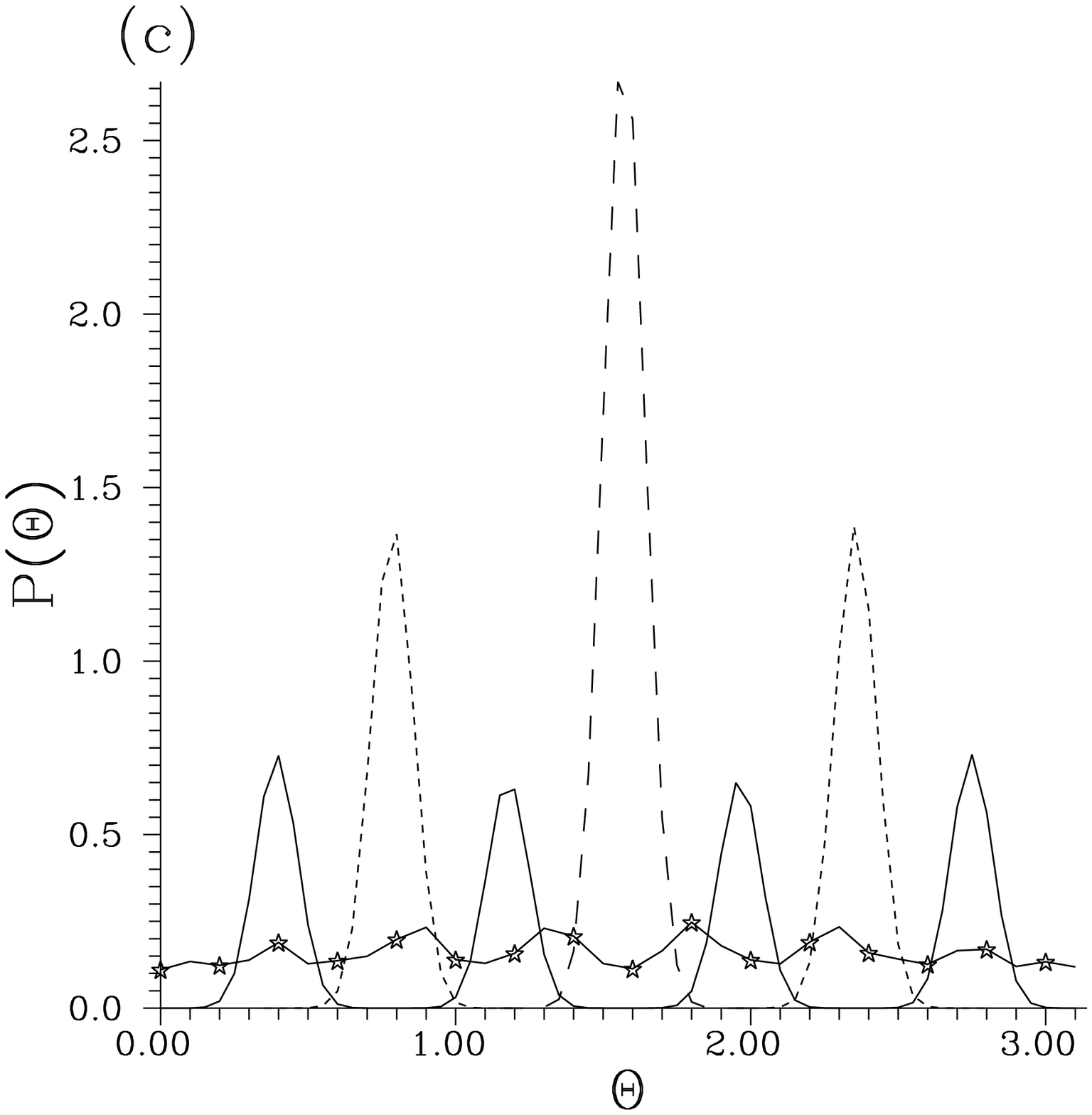} } \vspace{.1cm} \caption{ The
phase distribution $P(\Theta)$ against $\Theta$. Figure (a) is
given for $(\alpha,\epsilon,k)=(3,0,1)$ with $(T,g)=(T_r,1)$
(solid curve), $(T_r/2,1)$ (short-dashed curve), $(T_r/4,1)$
(long-dashed curve), $(19.47003,0.5)$ (star-centered curve)  and
$(10.18,0.5)$ (bell-centered curve).
 For (b) $(\alpha,\epsilon,m,g,k)=(7,1,0,0.5,1)$ with
$T=40.99995$ (second-revival time, solid curve), $9.099998$
(collapse-time, short-dashed curve) and $21.00004$
(first-revival-time, long-dashed curve). In (b) the star-centered
curve is given for
$(\alpha,\epsilon,m,g,k,T)=(7,0,1,0.5,1,41.04966)$. For (c)
$(\alpha,\epsilon,m,g,k)=(7,1,0,1,2)$
 for $T=\pi/4$ (solid curve),
$\pi/2$ (short-dashed curve) and $\pi$ (long-dashed curve). In (c)
star-centered curve is given for
$(\alpha,\epsilon,m,g,k,T)=(7,0,0,1,3,\pi/4)$.}
\end{figure}
%%%%%%%%%%%%%%%%%%%%%%%%%%%%%%%%%%%%%%%%%%%%%%%%%%%%%%%%%%%%%%%%%

\begin{eqnarray}
\fl
\begin{array}{lr}
P(\Theta) =\frac{1}{2\pi}\Bigl|
\sum\limits_{n,n'=0}^{\infty}\hat{\rho}_{n,n'}(T)\exp(i\Theta)\Bigr|^2
\\
\\
= \frac{1}{2\pi}\Bigl\{1+2
\sum\limits_{n>n'}^{\infty}C(n,m)C(n',m) [
X_1(n,T)X_1(n',T)+X_2(n,T)X_2(n',T)
\\
\\
+ X_3(n,T)X_3(n',T) +X_4(n,T)X_4(n',T)]\cos[(n-n')\Theta]\Bigr\},
\label{14p}
\end{array}
\end{eqnarray}
where we have considered that the phase reference $\Theta_0=0$. It
is evident that $P(\Theta)=P(-\Theta)$. As a result of this
symmetry we have plotted $P(\Theta)$ against $\Theta$ for $0\leq
\Theta\leq \pi$  in Figs. 4(a)--(c) for given values of the
interaction parameters.
 From Fig. 4(a), i.e. for the symmetric case, one can see that  in the
course of the revival time the $P(\Theta)$ exhibits single-peak
structure at $\Theta=0$. This behavior is close to that of the
initial coherent light. Nevertheless, at one-half (quarter) of the
revival time two small lateral  peaks (wings) occur. This
indicates  generation of the asymmetric cat states in the system.
 This behavior is not clear in the
corresponding $W$ function (see Figs. 2). It seems that the
non-occurrence of the lateral peaks in the $W$ function is related
to that $\alpha$ is considerably small. In this case the
contribution of the different components in the $W$ function will
be close to the phase-space origin and destructively  interfere
with each others. Furthermore, from the star-centered and
bell-centered curves in Fig. 4(a) (, i.e. asymmetric case)
$P(\Theta)$ exhibits two-peak (four-peak) structure through the
revival (collapse) time. In other words, two-component and
four-component cat states can be generated in the asymmetric TJCM
based on the values of the interaction parameters. This has been
explained for the $W$ function as a generation of microscopic cat
states.  Information on the interference in phase space is
 shown in Fig. 4(b) for the asymmetric case. For the asymmetric case
there are two forms of the  revival patterns: one is smooth and
the other is asymmetric. Thus in Fig. 4(b) we have plotted the
$P(\Theta)$ for the first  and second revival times. As is well
known that for $\Theta_0=0$  the $P(\Theta)$ of the cat state
possesses single-peak at $\Theta=0$ and two wings as
$\Theta\rightarrow \pm \pi$ \cite{phase}.
 This form  changes by switching on the
interaction where in the course of the collapse time asymmetric
eight peaks are created, which are transformed to
 symmetric four peaks  at the first revival time and eventually
are collapsed to two-peak structure  at the second revival time.
It is obvious  that the shape of the peaks in $P(\Theta)$ includes
information on the shape of the revival patterns, i.e. when the
$P(\Theta)$ exhibits smooth  peaks the revival patterns are smooth
and vice versa. Furthermore, the deformation in the $P(n)$ leads
to deformation in the peaks  of $P(\Theta)$ (see the star-centered
curve in Fig. 4(b)). Also we have noted that when the values of
the ratio $g$ changes the number of peaks in $P(\Theta)$ changes,
too. Furthermore, we have noted that the behavior of the case
$k=2$ is similar to that of $k=1$, however, for the former  the
peaks are always smooth and narrower than those for the latter
(see Fig. 4(c)). Also when $k>2$ the $P(\Theta)$ exhibits
multipeak structure (see the star-centered curve in Fig. 4(c)).
This situation is similar to that of the two-mode JCM \cite{tpha}.

%%%%%%%%%%%%%%%%%%%%%%%%%%%%%%%%%%%%%%%%%%%%%%%%%%%%%%%%%%%%%
\section{Entanglement}
%%%%%%%%%%%%%%%%%%%%%%%%%%%%%%%%%%%%%%%%%%%%%%%%%%%%%%%%%%%%%%
Entanglement is at the heart of quantum information theory.
Various efforts are done to characterize qualitatively and
quantitatively the entanglement properties of the quantum systems.
This has been motivated by the progress in the experimental
techniques aiming to create entangled states, which provide
phenomena very different from the classical physics \cite{zly}. In
this section we investigate the the entanglement for the system
under consideration using the tangle $I$ defined in  \cite{rung}.
 Based on the symmetry exchange for the system we
investigate two forms of tangles, which are field-atoms and
one-atom-remainder. In other words, assume that $f, A_1$ and $A_2$
denoting the field, first atom and second atom, respectively.
Therefore, the field-atoms tangle  ($f-A_1A_2$) and
one-atom-remainder tangle ($A_1-fA_2$), say, are define as
\cite{tess}:
\begin{eqnarray}
\fl
\begin{array}{lr}
I_{f-A_1A_2}(T)=2[1-{\rm Tr}\hat{\rho}^2_f(T)]=2[1-{\rm
Tr}\hat{\rho}^2_{A_1A_2}(T)],
\\
\\
I_{A_1-fA_2}(T)=2[1-{\rm Tr}\hat{\rho}^2_{A_1}(T)]=2[1-{\rm
Tr}\hat{\rho}^2_{fA_2}(T)], \label{en1}
\end{array}
\end{eqnarray}
where, e.g., $\hat{\rho}_f(T)$ is the reduced density matrix of
the field, which can be obtained by tracing the total density
matrix of the system over a complete set of the atoms $A_1A_2$.
The forms (\ref{en1}) quantify the degree of entanglement to which
the ensembles behave as a collective entity. It is worth
mentioning that the tangle here includes the notion of the purity.
Generally, when $I_{f-A_1A_2}(T)=0$, say, the parties $f$ and
$A_1A_2$  are completely disentangled, but of course could be in
states different from those of the initial ones. Nevertheless,
when $I_{f-A_1A_2}(T)=2$ the parties  are maximally entangled.
 In the following we use the
terminologies:  $I_{f-A_1A_2}(T)\leq 1$ for weak entangled parties
and $I_{f-A_1A_2}(T)>1$ for strong entangled parties.
%%%%%%%%%%%%%%%%%%%%%%%%%%%%%%%%%%%%%%%%%%%%%%%%%%%%%%%%%%%%%%%
\begin{figure} \vspace{0cm}
\centerline{\epsfxsize=6cm \epsfbox{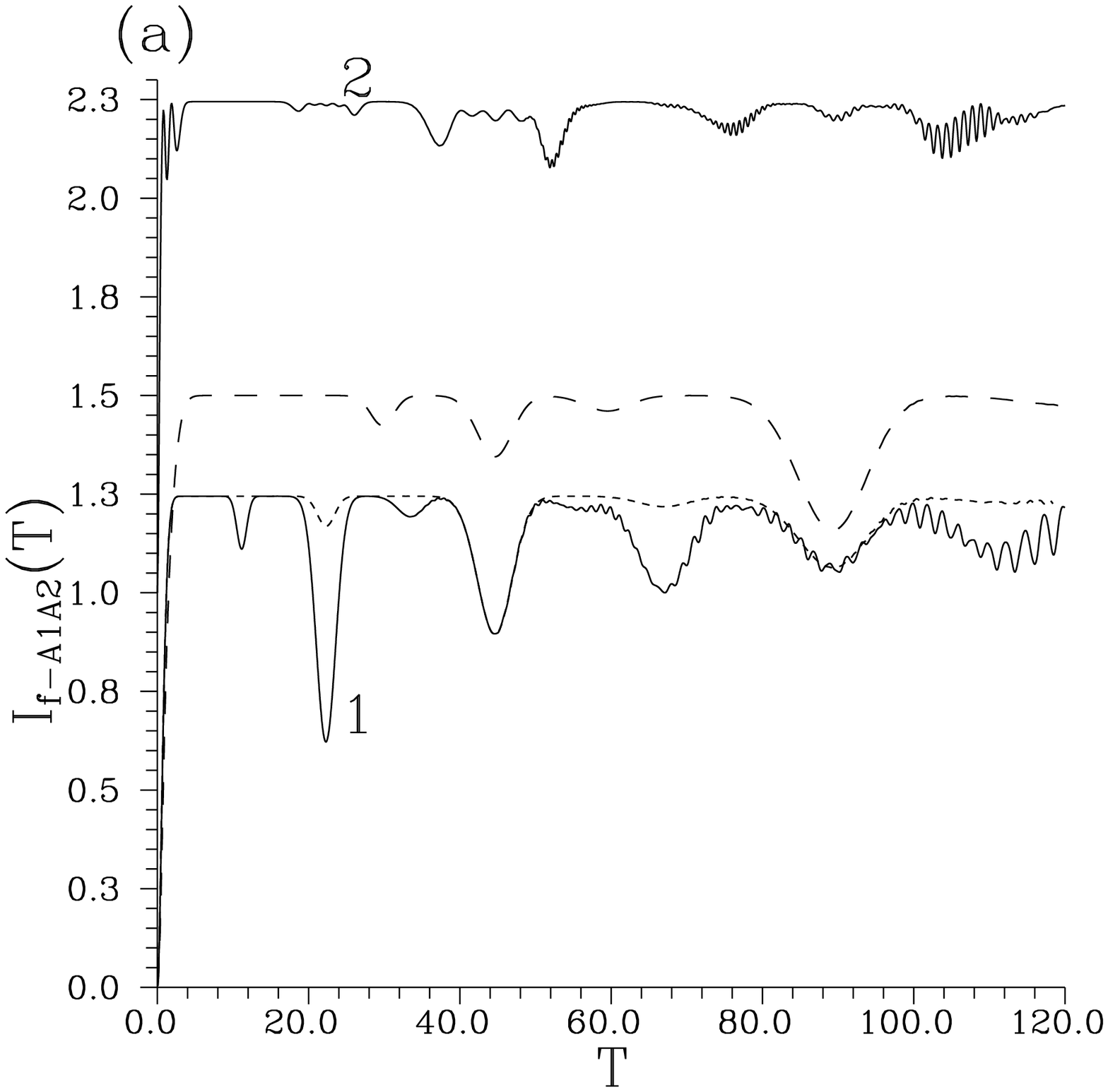}
\\\\\\\\\\\
\epsfxsize=5.5cm \epsfbox{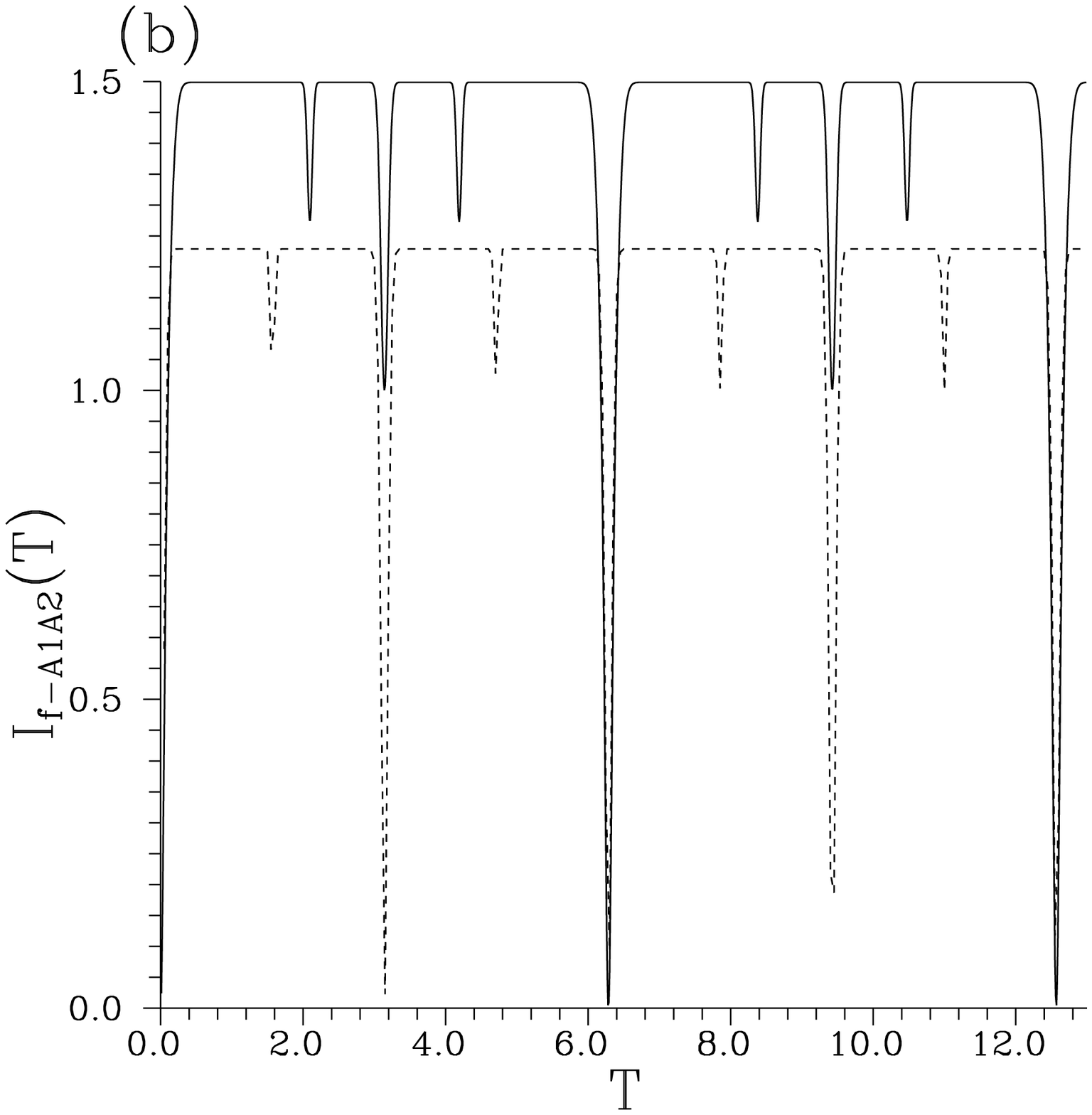}
\\\\\\\\\\\
\epsfxsize=5.5cm \epsfbox{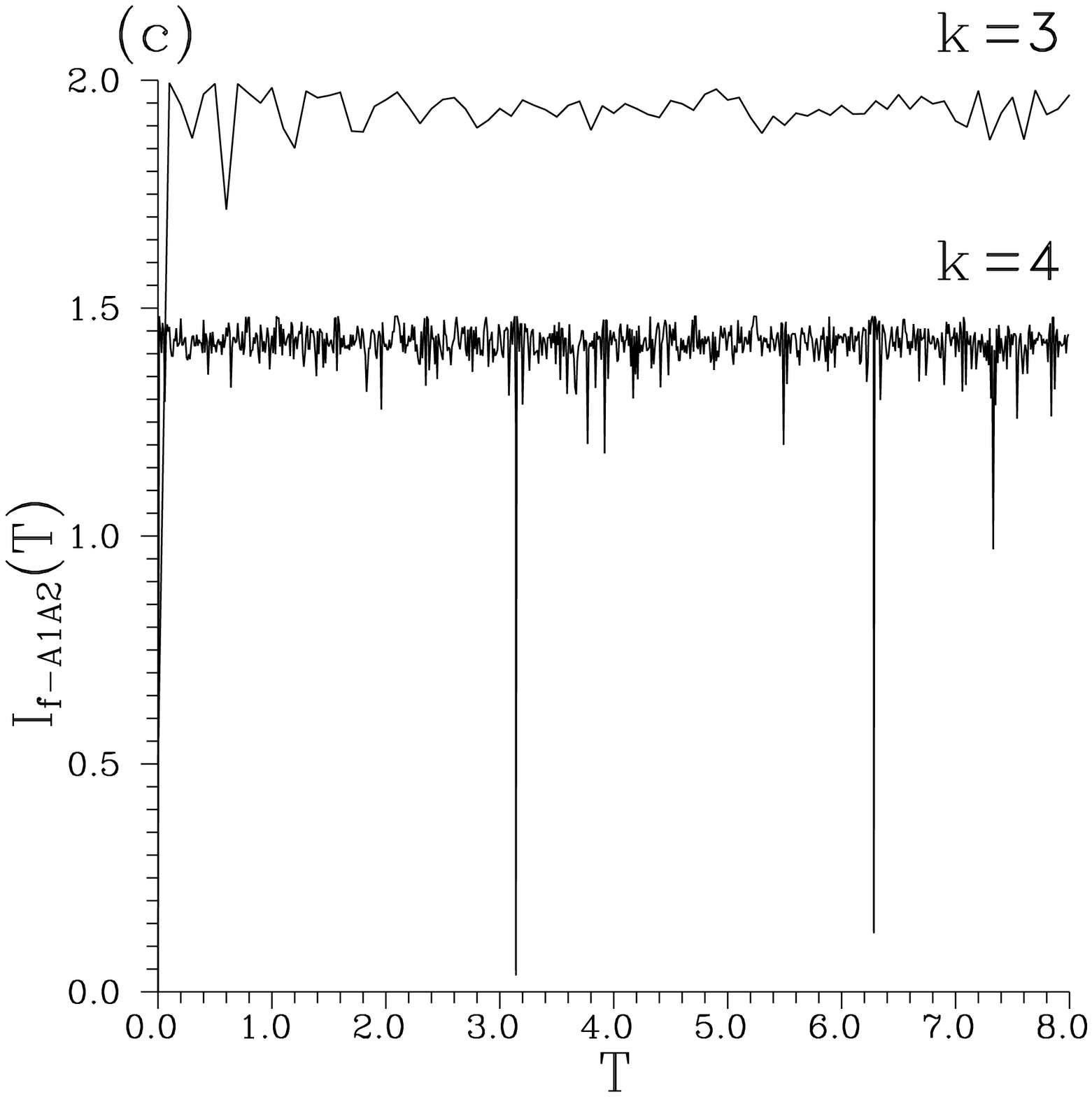}
 } \vspace{.1cm} \caption{ The tangle
 $I_{f-A_1A_2}(T)$ against the scaled time $T$
when $\alpha=7$ for (a) $(\epsilon,g,m,k)=(0,1,0,1)$ (the
short-dashed curve), $(0,0.5,0,1)$ (the long-dashed curve),
$(1,1,0,1)$ (the solid-curve-1) and $(0,0.5,2,1)$ (the solid-curve
$2$); (b) $(\epsilon,g,m,k)=(0,1,0,2)$ (short-dashed curve) and
$(0,0.5,0,2)$ (solid curve); and (c) $(\epsilon,g,m)=(0,1,0)$. The
solid-curve-2 in (a) is shifted by $1$. }
\end{figure}
%%%%%%%%%%%%%%%%%%%%%%%%%%%%%%%%%%%%%%%%%%%%%%%%%%%%%%%%%%%%%%%%%

%%%%%%%%%%%%%%%%%%%%%%%%%%%%%%%%%%%%%
\subsection{Field-atoms tangle}
%%%%%%%%%%%%%%%%%%%%%%%%%%%%%%%%%%%%%%
In this part we study the dynamical evolution for the tangle
$I_{f-A_1A_2}(T)$ for the system under consideration, which can be
easily evaluated as
\begin{eqnarray}
\fl
\begin{array}{lr}
I_{f-A_1A_2}(T)=2-2 \sum\limits_{n,n'=0}^{\infty} \Bigl\{
C(n,m)C(n',m) X_1(n,T)X_1(n',T) +C(n-k,m)C(n'-k,m)
\\
\\
\times [X_2(n-k,T)X_2(n'-k,T) + X_3(n-k,T)X_3(n'-k,T)]
\\
\\
+ C(n-2k,m)C(n'-2k,m) X_4(n-2k,T)X_4(n'-2k,T)  \Bigr\}^2.
\label{en2}
\end{array}
\end{eqnarray}
We have plotted (\ref{en2}) in Figs. 5(a)--(c) for different
values of the interaction parameters. For all curves one can
observe that at $T=0$, where the bipartite is disentangled,
$I_{f-A_1A_2}(T)=0$. We start with the symmetric case in  Fig.
5(a). From the short-dashed curve it is obvious that
 when $t>0$ the entanglement is established
between the two parties and $I_{f-A_1A_2}(T)$ goes rapidly to the
strong-entangled region, stays for while, dips to weak entangled
region and eventually becomes  steady for large interaction time.
This behavior is completely different from that of the purity of
the JCM, which exhibits oscillatory behavior in a good
correspondence with the revival patterns in the atomic inversion,
and reduces to a pure state at the middle of the collapse time
\cite{fais}. From the solid-curve-1 one can observe that the
interference in phase space
 decreases the degree of entanglement at particular
values of the interaction time and increases the oscillatory
behavior in the tangle (compare the short-dashed curve and
solid-curve-1). For the asymmetric case, i.e. the long-dashed
curve and solid-curve-2, one can observe generally that the degree
of entanglement becomes greater than that of the symmetric case
(compare short-dashed and long-dashed curves). Also the influence
of including  Fock state in the optical cavity on the behavior of
the $I_{f-A_1A_2}(T)$ is shown by the solid-curve-2. From this
curve it is obvious  that  the oscillatory behavior in
$I_{f-A_1A_2}(T)$ is  increased compared to that of the initial
coherent light, however, this behavior is not consistent with the
RCP in the corresponding atomic inversion (see Fig. 1(b)). Now we
draw the attention to Fig. 5(b) in which we have plotted
$I_{f-A_1A_2}(T)$ for the  two-photon transition case, i.e. $k=2$.
This figure indicates that the bipartite can be periodically
disentangled for the symmetric and asymmetric cases. For the
symmetric case the $I_{f-A_1A_2}(T)$ stays in the strong-entangled
region most of the interaction time and goes to zero periodically
with period $\pi$. This can be easily understood, e.g. for
$(\epsilon,g,m)=(0,1,0)$, in the framework of the SR as follows.
In this regime such terms $1/n,1/n^2,..$ tend to zeros leading to
that $\zeta_n\simeq 2(n+5/2)$ (cf. (\ref{I4i})). Thus the state
(\ref{10}) can be expressed in the following asymptotic form:
%%%%%%%%%%%%%%%%%%%%%%%%%%%%%%%%%%%%%%%%%%%%%%%%%%%%%%%%%%%%%%%
\begin{figure} \vspace{0cm}
\centerline{\epsfxsize=7cm \epsfbox{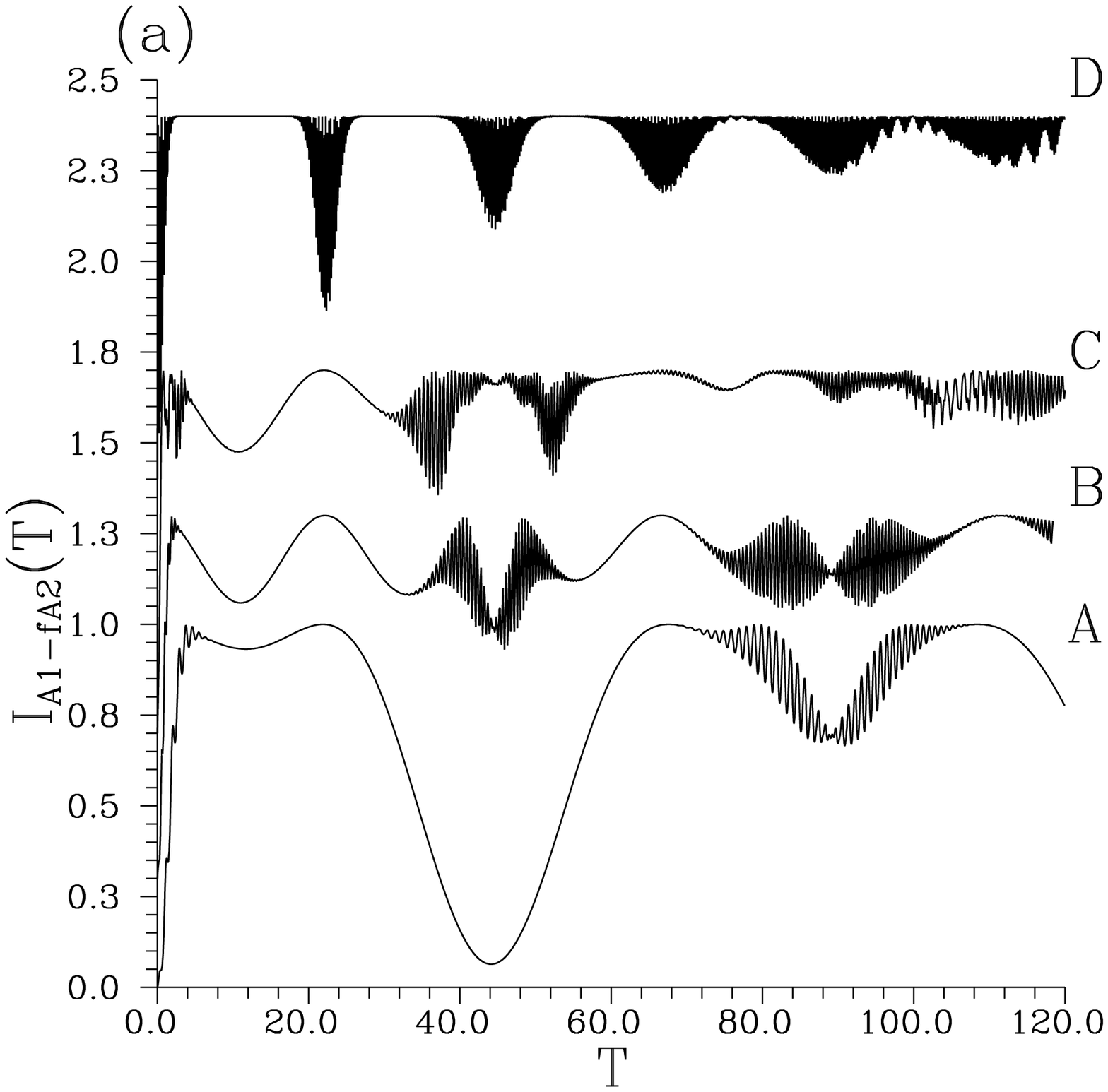}
\\\\\\\\\\\
\epsfxsize=7cm \epsfbox{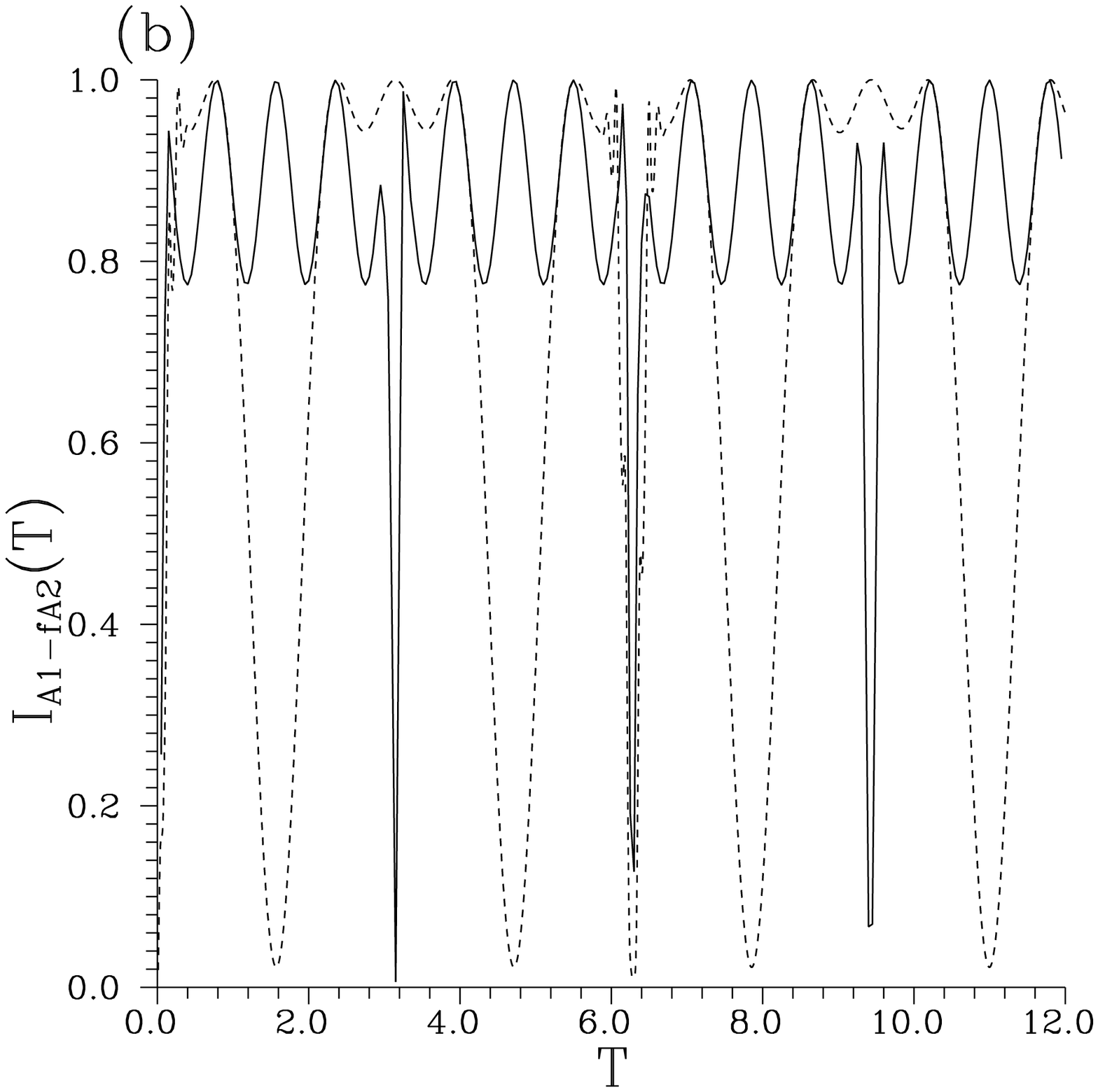}
 } \vspace{.1cm} \caption{ The
 $I_{A_1-fA_2}(T)$ against the scaled time $T$ when
$\alpha=7$ for  (a) $(\epsilon,g,m,k)= (0,0.5,0,1)$ (curve A),
$(0,1,0,1)$ (curve B), $(0,0.5,2,1)$ (curve C) and
  $(1,1,0,1)$ (curve D), and (b) $(\epsilon,g,m,k)= (0,1,0,2)$ (solid curve) and
$(0,0.5,0,2)$  (dashed-curve). In (a) the curves A--D are shifted
from bottom by $0,0.3,0.7,1.4$, respectively.}
\end{figure}
%%%%%%%%%%%%%%%%%%%%%%%%%%%%%%%%%%%%%%%%%%%%%%%%%%%%%%%%%%%%%%%%%
\begin{eqnarray}
\begin{array}{lr}
|\Psi(T)\rangle\simeq
\frac{1}{2}\sum\limits_{n=0}^{\infty}C(n,0)\Bigl\{[\cos(2nT+5T)+1]|+,+,n\rangle
-2i \sin(2nT+5T)
\\
\\
\times [|+,-,n+4\rangle+|-,+,n+2\rangle]
+[\cos(2nT+5T)-1]|-,-,n+4\rangle \Bigr\}. \label{10as}
\end{array}
\end{eqnarray}
When  $T=s\pi$, $s$ is odd number, the state (\ref{10as}) reduces
to
\begin{equation}
|\Psi(T)\rangle\simeq
\sum\limits_{n=0}^{\infty}C(n,0)|n+4\rangle\otimes \exp(i\pi)
|-,-\rangle . \label{11as}
\end{equation}
It is evident that in (\ref{11as}) the state of the radiation
field is quite similar to that of the initial one but with
four-photon shift. This behavior  is similar to that of the purity
of the JCM, but the forms of the phase factors and the number of
shifted photons in the two quantities are different (compare
(\ref{11as}) with (22)--(23) in \cite{fais}). Moreover, when
$T=s'\pi$ where $s'$ is even integer, (\ref{10as}) reduces to its
initial form. Also at $T=\pi s''/2$, $s''$ is odd integer, the
state (\ref{10as}) takes the form
\begin{eqnarray}
\begin{array}{lr}
|\Psi(T)\rangle\simeq
\frac{1}{2}\sum\limits_{n=0}^{\infty}C(n,0)\Bigl\{[
|+,+,n\rangle-|-,-,n+4\rangle
\\
\\
-2i (-1)^n
[|+,-,n+2\rangle+|-,+,n+2\rangle]
 \Bigr\}. \label{12as}
\end{array}
\end{eqnarray}
For (\ref{12as}) one can easily prove that $I_{f-A_1A_2}(T)=1$.
All these
 analytical facts are remarkable in Fig. 5(b). For the asymmetric
case, i.e. the solid curve in Fig. 5(b), we obtain behavior as
that of the symmetric case but the degree of entanglement is
greater and the disentanglement period is $2\pi$. Nevertheless,
for this case it is difficult to obtain explicit forms for the
state vector at $I_{f-A_1A_2}(T)=0$. It is obvious that
 the disentanglement period depends on the value of the
ratio $g$. Also  the behavior of the $I_{f-A_1A_2}(T)$ is
 consistent with the occurrence  of the RCP in the evolution of
the corresponding $\langle\sigma^{z}(T)\rangle$, we have checked
this fact. From Fig. 5(c) one can see that for
 $k=3$ the bipartite are maximally entangled, however, for $k=4$ the
 degree of entanglement is decreased and the bipartite becomes
periodically disentangled
 with period $\pi$.
This behavior is similar that of the purity of the JCM. Finally,
in Fig. 5(c)
 we have plotted $I_{f-A_1A_2}(T)$ of the asymmetric case only
 where that of the symmetric case provides quite similar behaviors.

%%%%%%%%%%%%%%%%%%%%%%%%%%%%%%%%%%%%%
\subsection{One-atom-remainder tangle}
%%%%%%%%%%%%%%%%%%%%%%%%%%%%%%%%%%%%%%
In this part we study the dynamical evolution for the tangle
$I_{A_1-fA_2}(T)$, which can be expressed as:
\begin{eqnarray}
\fl
\begin{array}{lr}
I_{A_1-fA_2}(T)=2-2 \sum\limits_{n,n'=0}^{\infty} \Bigl\{
|C(n,m)C(n',m)|^2[ |X_1(n,T)|^2+|X_3(n,T)|^2] [
|X_1(n',T)|^2+|X_3(n',T)|^2]
\\
\\
+|C(n,m)C(n',m)|^2[ |X_2(n,T)|^2+|X_4(n,T)|^2][
|X_2(n',T)|^2+|X_4(n',T)|^2]
\\
\\
 +2 C(n+k,m)C(n,m)C(n'+k,m)C(n',m)
[X_1(n+k,T)X^*_2(n,T)+X_3(n+k,T)X_4(n,T)]
\\
\\
\times [X_1(n'+k,T)X^*_2(n',T)+X_3(n'+k,T)X_4(n',T)] \Bigr\}.
\label{en3}
\end{array}
\end{eqnarray}
We have plotted $I_{A_1-fA_2}(T)$ in Figs. 6 for the given values
of the interaction parameters. The obvious remark from these
figures is that $0\leq I_{A_1-fA_2}(T)\leq 1$. This is completely
different from the behavior of the $I_{f-A_1A_2}(T)$ (compare
Figs. 5 and 6 as well as expressions (\ref{en2}) and (\ref{en3})).
This indicates that the degree of entanglement for the
one-atom-remainder tangle is less than that of the field-atoms tangle.
Also this shows  that the quantum entanglement cannot be equally
distributed  among many different objects. This  can be explained
as follows. We have an isolated system, i.e. the interaction with
the environment is neglected, and hence the energy in the field
and the two-atom system is periodically exchanged. In other words,
when the $k$ photons are annihilated from the radiation field they
are created, i.e. equally distributed, in
 the two-atom party and vice versa. This means that the energy
included in the $fA_2$ party is more than that in the $A_1$ party.
Actually,  the entanglement is a direct consequence of the energy
follow  between the two parties. The rate of follow of energy in
the bipartite of the $I_{f-A_1A_2}(T)$ is greater than that in the
$I_{A_1-fA_2}(T)$ and this could be the origin in the
 difference between the evolution of the two
tangles. Now from the curves B and D in Fig. 6(a), i.e. symmetric
case, one can observe that in contrast to $I_{f-A_1A_2}(T)$ the
$I_{A_1-fA_2}(T)$ exhibits an oscillatory behavior in a good
correspondence with the revival patterns in
$\langle\sigma_{z}(T)\rangle$. Also the interference in phase
space
 decreases the degree of entanglement in the
 $I_{A_1-fA_2}(T)$ (compare curves B and D).
 It is worth mentioning that the curve D is quite similar to that of
 the purity of the JCM (see Fig. 2 in \cite{rekd}).
 For the asymmetric case, and from the curve A
 one can observe that
$I_{A_1-fA_2}(T)$ exhibits compound behavior, i.e. it includes
periodic and oscillatory behaviors. Also the bipartite can be
approximately disentangled in the course of the first revival
pattern. As in $I_{f-A_1A_2}(T)$ the Fock state increases
 the range of the
oscillatory behavior in the $I_{A_1-fA_2}(T)$ (see
 the curve C).
 From
 Fig. 6(b), which is given for the case
$k=2$, one can see that for the symmetric case the
$I_{A_1-fA_2}(T)$ exhibits  disentanglement periodically
 with period $\pi$ (cf. (\ref{11as})) as in $I_{f-A_1A_2}(T)$
 and also includes oscillatory behavior.
 The $I_{A_1-fA_2}(T)$ of the asymmetric case (, i.e. dashed curve)
provides two types of disentanglement long lived (at $T=\pi/2,
3\pi/2$) and
 instantaneous (at $T= 2\pi$) and they occur periodically with
 period $2\pi$. For the
  cases $k=3,4$ we have noted that the $I_{A_1-fA_2}(T)\simeq 1$
  with periodic disentanglement
only for the case $k=4$.

%%%%%%%%%%%%%%%%%%%%%%%%%%%%%%%%%%%%%%%%%%%%%%%%%%%%%%%%%%%%%
\section{Conclusion }
%%%%%%%%%%%%%%%%%%%%%%%%%%%%%%%%%%%%%%%%%%%%%%%%%%%%%%%%%%%%%%
In this paper we have treated the system of two two-level atoms
interacting with multiphoton single-mode field. The two atoms and
the field are initially prepared  in the excited atomic states and
in the superposition of  displaced number states, respectively. We
have investigated the behavior of the atomic inversion, $W$
function, phase distribution and entanglement. We have shown that
the TJCM can generate asymmetric (symmetric) cat states at quarter
of the
 revival time for the symmetric (asymmetric) case.  Also this has been confirmed
 in the behavior of the phase distribution. Moreover, we have
 deduced the asymptotic form for the $W$ function when $\alpha>>1$.
 Also we have noted that when the values of
 $g$ changes the number of the components of the
generated cat state in the system changes, too. The $P(\Theta)$
exhibits multipeak structure for $k>2$.
 For the entanglement we have investigated two types, namely, field-atoms and
one-atom-remainder tangles. We have shown that the degree of
entanglement in $I_{f-A_1A_2}$ is much  greater than that in
 $I_{A_1-fA_2}$. We have explained this in the framework of the rate of
  energy follow between
  different parties of the system.
We have obtained the following facts related to entanglement. The
degree of entanglement for the asymmetric case is greater than
that of the symmetric one. The interference in phase space
 decreases the degree of entanglement.
 There is a similarity between the dynamical behavior of the
$I_{f-A_1A_2}(T)$ and the purity of the JCM only for $k>1$.

\ack I would like to thank the Abdus Salam International Centre
for Theoretical Physiscs, Strada Costiers, 11 34014 Trieste Italy
for the hospitality and financial support under the system of
associateship, where a part of this work is done.

%%%%%%%%%%%%%%%%%%%%%%%%%%%%%
\section*{References}
%%%%%%%%%%%%%%%%%%%%%%%%%%%%%%

\end{document}